\documentclass[12pt,a4paper]{article}
\usepackage{graphicx}
\usepackage{amssymb}
\usepackage{amsmath}
\numberwithin{equation}{section} 

\usepackage{amsfonts}
\usepackage{bm}
\usepackage[version=4]{mhchem} 
\usepackage{color}
\usepackage{cancel}
\usepackage[vcentermath]{youngtab}
\usepackage{braket}

\usepackage[utf8]{inputenc} 

\usepackage{url} 

\usepackage{appendix} 
\usepackage{booktabs} 
\usepackage{tablefootnote} 
\usepackage{multirow} 
\usepackage[inline]{enumitem} 
\usepackage{float} 
\usepackage{caption} 
\usepackage{subcaption} 

\definecolor{Orange}{cmyk}{0,0.61,0.87,0}
\definecolor{JungleGreen}{cmyk}{0.99,0,0.52,0}
\definecolor{OliveGreen}{cmyk}{0.64,0,0.95,0.40}
\definecolor{Brown}{cmyk}{0,0.81,1,0.60}
\definecolor{RoyalBlue}{cmyk}{0.71,0.53,0,0.12}

\usepackage[colorlinks=true, linkcolor=OliveGreen, citecolor=RoyalBlue,
urlcolor=Brown]{hyperref}

\newcommand{\hc}{\text{h.c.}}
\newcommand{\SU}{\text{SU}}
\newcommand{\Sp}{\text{Sp}}
\newcommand{\SO}{\text{SO}}
\newcommand{\U}{\text{U}}
\newcommand{\dd}{\text{d}}
\newcommand{\Lag}{\mathcal{L}}
\newcommand{\id}{\mathbb{I}}
\newcommand{\tr}{\,\text{Tr}\,}

\newcommand{\rep}[1]{\textbf{#1}}
\newcommand{\brep}[1]{{\bf \overline{#1}}}
\newcommand{\e}{\textbf{1}}

\newcommand{\bchi}{\bar{\chi}}
\newcommand{\bpsi}{\bar{\psi}}
\newcommand{\tPhi}{\widetilde{\Phi}}

\newcommand{\tDelta}{\widetilde{\Delta}}

\newcommand{\x}{\,\,\,}

\usepackage{array}   
\newcolumntype{C}{>{$}c<{$}} 
\usepackage{slashed}

\newcommand{\icol}[1]{
  \left(\begin{smallmatrix}#1\end{smallmatrix}\right)%
}

\newcommand*\xbar[1]{%
 \kern0.5ex%
  \hbox{%
   \kern0.2ex%
      \vbox{%
      \hrule height 0.5pt 
      \kern0.5ex
      \hbox{%
        \kern-0.1em
        \ensuremath{#1}%
        \kern-0.1em
      }%
    }%
  }%
}

\newcommand{\gsim}{\lower.7ex\hbox{$
\;\stackrel{\textstyle>}{\sim}\;$}}
\newcommand{\lsim}{\lower.7ex\hbox{$
\;\stackrel{\textstyle<}{\sim}\;$}}

\makeatletter
\newsavebox\myboxA
\newsavebox\myboxB
\newlength\mylenA

\newcommand*\xoverline[2][0.75]{%
    \sbox{\myboxA}{$\m@th#2$}%
    \setbox\myboxB\null
    \ht\myboxB=\ht\myboxA%
    \dp\myboxB=\dp\myboxA%
    \wd\myboxB=#1\wd\myboxA
    \sbox\myboxB{$\m@th\overline{\copy\myboxB}$}
    \setlength\mylenA{\the\wd\myboxA}
    \addtolength\mylenA{-\the\wd\myboxB}%
    \ifdim\wd\myboxB<\wd\myboxA%
        \rlap{\hskip 0.5\mylenA\usebox\myboxB}{\usebox\myboxA}%
    \else
        \hskip -0.5\mylenA\rlap{\usebox\myboxA}{\hskip 0.5\mylenA\usebox\myboxB}%
    \fi}
\makeatother
\begin{document}
\begin{titlepage}

\begin{flushright}
UMN--TH--3923/20
\end{flushright}

\vskip 1.35cm
\begin{center}

{\Large{\bf A Composite Higgs with a Heavy Composite Axion}}

\vskip 1.5cm
{\bf Tony Gherghetta, Minh D. Nguyen}

\vskip 0.5cm

{{\it School of Physics and Astronomy,\\ University of Minnesota,\\
 Minneapolis, Minnesota 55455, USA}\\[3pt]}

\date{\today}

\vskip 1.5cm
\begin{abstract}

We consider the strong dynamics associated with a composite Higgs model that simultaneously produces dynamical axions and solves the strong CP problem.
The strong dynamics arises from a new $\Sp$ or $\SU(4)$ hypercolor gauge group containing QCD colored hyperfermions that confines at a high scale. The hypercolor global symmetry is weakly gauged by the Standard Model electroweak gauge group and an enlarged color group, $\SU(N+3)\times \SU(N)'$.
When hyperfermion condensates form, they not only lead to an $\SU(5)/\SO(5)$ composite Higgs model but also spontaneously break the enlarged color group to $\SU(3)_c\times \SU(N)_D$. At lower energies, the  $\SU(N)_D$ group confines, producing two dynamical axions that eliminates all CP violation. Furthermore, small instantons from the $\SU(N)'$ group can enhance the axion mass, giving rise to TeV scale axion masses that can be detected at collider experiments. Our model provides a way to unify the composite Higgs with dynamical axions, without introducing new elementary scalar fields, while also extending the range of axion masses that addresses the strong CP problem.

\end{abstract}

\end{center}
\end{titlepage}

\section{Introduction}

The axion is a popular solution to the strong CP problem. It arises as a Nambu-Goldstone boson from an anomalous $\U(1)$ Peccei-Quinn (PQ) global symmetry~\cite{Peccei:1977hh, Peccei:1977ur,  Weinberg:1977ma, Wilczek:1977pj}, that is spontaneously broken at some high scale $f_{\rm PQ}\simeq 10^9 - 10^{12}$ GeV. A crucial aspect is that the global symmetry is explicitly broken by non-perturbative QCD dynamics which generates an axion potential. At the potential minimum, the axion precisely cancels the $\bar\theta$ term in QCD, and therefore provides a dynamical reason for why there is no strong CP violation. The axion mass is then uniquely fixed up to the PQ symmetry breaking scale to be of order $10^{-6}-10^{-3}$ eV~\cite{diCortona:2015ldu}. Even though its couplings are suppressed by $f_{\rm PQ}$, this has led to an extensive search for the so-called invisible axion.

Given the active experimental effort it behooves us to consider possible ways to vary the axion mass and couplings that can arise from UV modifications of QCD. A number of proposals were previously considered in Refs.~\cite{Holdom:1982ex, Dine:1986bg, Flynn:1987rs}, and more recently in Refs.~\cite{Agrawal:2017ksf, Fuentes-Martin:2019bue, Csaki:2019vte, Gherghetta:2020keg}. 
A particularly interesting possibility is that associated with enlarging the color gauge group, where new nonperturbative contributions to the axion mass are generated~\cite{Gherghetta:2016fhp, Gaillard:2018xgk}. We assume that the QCD gauge group, $\SU(3)_c$ is embedded into $\SU(N+3)\times \SU(N)'$ with $N\geq 3$.
Although this enlarged color group can simply be broken with a new elementary Higgs sector~\cite{Gaillard:2018xgk}, a potentially richer framework is to cause the breaking with new strong dynamics. Besides the advantage of eliminating fundamental scalar fields, this approach offers the intriguing possibility that the same strong dynamics is also responsible for a composite Higgs. In fact, the strong dynamics associated with an enlarged color group can provide all the ingredients for a composite Higgs with partial compositeness as well as a dynamical axion with new contributions to the axion mass.

While composite Higgs models have been extensively explored as an effective low energy description based on the AdS/CFT correspondence (for a review, see ~\cite{Panico:2015jxa}), we will instead consider the UV completions of composite Higgs models given in Refs.~\cite{Barnard:2013zea, Ferretti:2013kya, Ferretti:2014qta}. These models contain colored fermion constituents that are needed to produce top partner bound states in order to implement the partial compositeness mechanism. An early approach that combined the axion and Higgs was considered in Ref.~\cite{Redi:2012ad}, although without an underlying four-fermion structure responsible for partial compositeness. In this paper we will extend the UV descriptions of composite Higgs models to incorporate the enlarged color breaking and generate dynamical axions. The new strong dynamics, referred to as {\it hypercolor}, will be based on the gauge groups $\Sp(2N_{\rm HC})$~\cite{Barnard:2013zea, Ferretti:2013kya}, where $N_{\rm HC}$ is a positive integer, and $\SU(4)_{\rm HC}$~\cite{Ferretti:2014qta}. With these gauge groups, the colored hyperfermions can transform in the fundamental representation of the hypercolor group. This restriction to the smallest representation preserves asymptotic freedom and leads to confinement of the hypercolor gauge group. The uncolored hyperfermions are chosen to be in the antisymmetric representation with an $\SU(5)$ global symmetry. Thus, there is a preference for $\SU(5)/\SO(5)$ composite Higgs models~\cite{Mrazek:2011iu, Ferretti:2014qta}.

When hypercolor becomes strong, two hyperfermion condensates will form. The first condensate, associated with the uncolored hyperfermions will give rise to Nambu-Goldstone bosons which contain the Standard Model Higgs doublet. As usual in composite Higgs models, the Standard Model electroweak symmetry is not broken by the hypercolor dynamics. A second condensate associated with colored hyperfermions will break the enlarged color group $\SU(N+3)\times \SU(N)'$ to $\SU(3)_c \times \SU(N)_D$. A gauged Nambu-Jona-Lasinio model can be used to find parameter regions where this breaking occurs. Below the hypercolor confinement scale, the fermion content of the $\SU(N)_D$ group will then allow confinement at a scale $\gtrsim$~10 TeV. The anomalous, axial $\U(1)$ symmetries associated with the global symmetries of the hypercolor and $\SU(N)_D$ dynamics plays the role of the PQ symmetry, producing several dynamical axions.   
By introducing massless fermions that transform under the respective PQ symmetries, all CP violation can be eliminated above the confinement scale, while the corresponding dynamical axions can eliminate all CP violation below the confinement scale. In particular, since some $\SU(N)_D$ massless fermions are charged under QCD, the corresponding dynamical axion will solve the strong CP problem. In fact the $\SU(N)_D$ dynamics below the hypercolor confinement scale is similar to the axicolor model of the invisible axion~\cite{Kim:1984pt, Choi:1985cb}.

The enlarged color group also allows dynamical axions to enhance their mass due to small instanton effects~\cite{Agrawal:2017ksf, Gaillard:2018xgk}. In particular, by assuming that the $\SU(N)'$ gauge coupling is sufficiently large, small instantons of size of order the hypercolor confinement scale (where the enlarged color group is broken) can give sizeable axion mass contributions. For example, assuming that $\SU(N)_D$ confines at 10 TeV, we obtain dynamical axion mass values near the TeV scale corresponding to a hypercolor confinement scale of approximately $1000$ TeV.
Alternatively, the hypercolor confinement scale can be near the GUT scale ($\simeq 10^{15}$~GeV) and also give rise to TeV scale axion masses. Furthermore, since the hypercolor confinement scale is well above the TeV scale, the composite Higgs sector will necessarily be tuned. In particular, if the hypercolor confinement scale is near $10^{10}$ GeV, it can also motivate why the Higgs quartic coupling vanishes at this scale. Since the Higgs is a Nambu-Goldstone boson, it naturally has a flat potential and provides the necessary boundary condition to radiatively generate the Higgs quartic coupling at the electroweak scale~\cite{Redi:2012ad}.

Our results suggest that broad searches for axion-like particles outside the usual invisible axion window, could in fact be relevant for the strong CP problem. For example, TeV scale axions can decay into photons and $Z$ bosons producing observable signals at the LHC. Furthermore, the strong dynamics associated with $\SU(N)_D$ can lead to colored bound states that may be accessible at future collider experiments.
In addition our model has several possible dark matter candidates, including an axion-like particle and a possible dark photon that could arise by gauging a $\U(1)$ global symmetry of the strong dynamics.
While TeV scale axions provide the most promising experimental signals there is also the possibility of detecting a colored octet fermion at the TeV scale. This state arises from considering the massless $\SU(N+3)$ fermion to be in a real representation, in contrast to a pseudoreal representation in which all the QCD colored fermions are confined at the $\SU(N)_D$ scale.

The outline of the paper is as follows.
In section \ref{sec:hc_enlarged_color} we introduce the hypercolor sector. Two possible hypercolor gauge groups that can incorporate an enlarged QCD color group are discussed in Section~\ref{subsec:hc_ptcl_content}. The possible global symmetry breaking by hyperfermion condensates is studied in Section \ref{subsec:hc_chi_potential} using a gauged Nambu-Jona-Lasinio model. The enlarged color symmetry breaking by the hyperfermion condensates is discussed in Section~\ref{subsec:Enlargedcolorbreaking}, and the embedding of the Standard Model quarks is presented in Section~\ref{subsec:SMquarkembedding}. In Section~\ref{sec:compositeaxion} we discuss the composite axion. We first consider a much simpler model without an enlarged color group, in Section~\ref{subsec:massless_fermions_hc}, that gives rise to the usual invisible axion. The enlarged color group is next considered in Section~\ref{subsec:massless_fermions_cut} where the anomalous PQ currents and corresponding axions are identified. The effect of small instantons on the axion mass from the $\SU(N)'$ group is considered in Section~\ref{subsec:smallinstantons}. The axion mass scales are calculated in Section~\ref{subsec:axionmass}, which are determined from the strong scale of the hypercolor group. We also give the conditions for asymptotic freedom of the strong gauge groups in Section~\ref{sec:RG_AsymptoticFreedom}. The phenomenological and cosmological aspects of the model are discussed in Section~\ref{sec:phenomenology}. Our concluding remarks are given in Section~\ref{sec:conclusion}. Further details of the one-loop effective potential are given in Appendix~\ref{app:A}, and the hypercolor Nambu-Goldstone boson spectrum are given in Appendix~\ref{app:B}.

\section{Hypercolor with Enlarged Color}
\label{sec:hc_enlarged_color}

An extended strong sector, where QCD is embedded into a larger gauge group such as $\SU(N+3)$, can solve the strong CP problem with a heavy axion~\cite{Gherghetta:2016fhp, Gaillard:2018xgk}.
Interestingly, composite Higgs models require top partners with colored constituents and these same constituents, charged under the hypercolor strong dynamics, can then play a role in breaking this enlarged color symmetry. A critical requirement in constructing such a model is that the hypercolor gauge coupling is asymptotically free and therefore hypercolor confines at low energies. This severely restricts the hyperfermion content of the model and typically favors colored constituents which are in the smallest representation of the gauge group. As will be shown in Section~\ref{sec:RG_AsymptoticFreedom}, the $\SU(5)$/$\SO(5)$ composite Higgs model is favored because the colored constituents can transform in the fundamental representation of the UV gauge group. Thus to obtain a composite Higgs with a heavy composite axion we will consider extensions of the $\SU(5)$/$\SO(5)$ model based on two choices of the hypercolor gauge group: the symplectic group $\Sp(2N_{\rm HC})$ with $N_{\rm HC}$ a positive integer, and  $\SU(4)_{\rm HC}$.

\subsection{Hypercolor Gauge Groups}
\label{subsec:hc_ptcl_content}

\subsubsection{\texorpdfstring{$\Sp(2N_{\rm HC})$}{Sp(2N)\_HC}}
\label{subsubsec:hc_symplectic}

We first consider the $\SU(5)/\SO(5)$ composite Higgs model based on the hypercolor gauge group $\Sp(2N_{\rm HC})$ with a traceless, antisymmetric hyperfermion $\psi^a$, transforming as a fundamental under an $\SU(5)$ global symmetry with index $a=1\dots 5$. 
The $\Sp(2N_{\rm HC})$ gauge invariant fermion bilinear is given by
\begin{equation}
    (\psi^a \psi^b) = \Omega_{ik} \Omega_{jl} \psi^a_{ij} \psi^b_{kl} \,,  
    \label{eq:hc_sp_psipsi}
\end{equation}
where $i, j = 1 \dots 2 N_{\rm HC} $ are $\Sp(2N_{\rm HC})$ indices, and $\Omega_{ij}$ is the symplectic metric
\begin{equation}
    \Omega_{ij} = 
    \begin{pmatrix}
        0            		& \id_{N_{\rm HC}} \\
        -\id_{N_{\rm HC}}   & 0
    \end{pmatrix} \,,
    \label{eq:hc_metric}
\end{equation}
with $\id_{N_{\rm HC}}$ the $N_{\rm HC} \times N_{\rm HC}$ identity matrix. The traceless condition can then be written as $\Omega_{ij}\psi^{ij} = 0$.
Besides describing a composite Higgs, the hypercolor dynamics must also produce colored, top partner bound states that can linearly mix with the elementary top quark. 
In addition to generating a large top Yukawa coupling, these states are required to explicitly break the global symmetry and generate a Higgs potential. 
This requires adding vector-like pairs of hyperfermions, $\chi,\bchi$ charged under $\SU(3)_c$ as first considered in Ref.~\cite{Barnard:2013zea}.

If the QCD gauge group is now extended to $\SU(N+3) \times \SU(N)'$, then the top partner hyperfermion content also needs to be enlarged to $\Psi_{\chi} = (\chi, \chi', \bchi, \bchi')$, where $\Psi_{\chi}$ are in the fundamental (pseudoreal) representation of $\Sp(2N_{\rm HC})$. 
As a result, the global symmetry group associated with $\Psi_{\chi}$ is $G_{\Sp} = \U(4N + 6)$. The minimal anomaly-free particle content is summarized in Table~\ref{tab:sp_particle_content}, where the electroweak gauge group $\SU(2)_L \times \U(1)_Y$ quantum numbers have been omitted. 
Typically, the electroweak group weakly gauges the custodial global symmetry $\SU(2)_L \times \SU(2)_R \times \U(1)_X \subset \SO(5)$~\cite{Ferretti:2013kya}.

\begin{table}[h]
    \centering
    \begin{tabular}{|l||c|c|c||c|c|}
    \cline{1-6}
                        & $\Sp(2N_{\rm HC})$    & $\SU(N+3)$        & $\SU(N)'$         &   $\SU(5)$                & $\U(4N + 6)$   \\ \hline
        $\psi$          & ${\bf A_2}$           & {\bf 1}           & {\bf 1}           &   {\bf 5}                 & {\bf 1}         \\ \hline
        $\chi$          & \multirow{4}{*}{\bf F}& {\bf F}           & {\bf 1}           &   \multirow{4}{*}{\bf 1}  & \multirow{4}{*}{\bf F} \\
        $\chi'$         &                       & {\bf 1}           & {\bf F}           &                           &                 \\ 
        $\bchi$         &                       & ${\bf \bar{F}}$   & {\bf 1}           &                           &                 \\
        $\bchi'$        &                       & {\bf 1}           & ${\bf \bar{F}}$   &                           &                 \\ 
    \cline{1-6}
    \end{tabular}
    \caption{The hypercolor fermion representations under the local symmetry, $\Sp(2N_{\rm HC})\times \SU(N+3) \times \SU(N)'$ and global symmetry, $\SU(5)\times \U(4N + 6)$.}
    \label{tab:sp_particle_content}
\end{table}

To investigate the global symmetry breaking pattern of the strong dynamics we construct a gauged NJL model~\cite{Nambu:1961tp}. For the composite Higgs sector this can be done by introducing four-fermion interactions consisting of $\psi^a$, and modifying the analysis in Ref.~\cite{Barnard:2013zea} to accommodate the $\SU(5)/\SO(5)$ coset. The leading interaction Lagrangian is given by
\begin{equation}
    \Lag_{\text{int}} = \frac{\kappa_{\psi}}{2N_{\rm HC}} (\psi^a \psi^b)(\psi^{\dagger}_a \psi^{\dagger}_b)\,.
    \label{eq:hc_sp_psi_int}
\end{equation}
Similarly to the NJL model, the vacuum structure of the $\Sp(2N_{\rm HC})$ strong dynamics can be studied via the introduction of a bilinear auxiliary field
\begin{equation}
    M^{ab} = - \frac{\kappa_{\psi}}{2N_{\rm HC}} (\psi^a \psi^b) \,,
    \label{eq:hc_sp_aux_m}
\end{equation}
where $\kappa_{\psi}$ is assumed to be real and positive. Since the representation of $\psi$ is real, $M^{ab}$ is symmetric and can be diagonalized via the Takagi diagonalization using an $\SU(5)$ global rotation of $\psi$. 
The effective potential of $M^{ab}$ admits a nontrivial, stationary fixed point where the singular values $m_{(i)}$ of $M^{ab}$ satisfy the mass gap equation
\begin{equation}
    \frac{1}{\zeta_{\psi}} \equiv \frac{4\pi^2}{\Lambda_{\rm UV}^2} \frac{1}{\kappa_{\psi}} =  1 - \frac{m_{(i)}^2}{\Lambda_{\rm UV}^2}  \log \left( 1 + \frac{\Lambda_{\rm UV}^2}{m_{(i)}^2} \right)\,, 
    \label{eq:gap}
\end{equation}
and $\Lambda_{\rm UV}$ represents a physical UV cutoff.  A nontrivial stable fixed point characterized by nonzero $m_{(i)}$ only exists if the dimensionless coupling parameter $\zeta_\psi$ is greater than one. This corresponds to large $\kappa_{\psi}$, i.e. the four-fermion coupling is strong. The scale $\Lambda_{\rm UV}$ can thus be interpreted as the confinement scale $\Lambda_{\rm HC}$ of $\Sp(2N_{\rm HC})$, where these couplings are generated. 
Thus for generic values of $\zeta_\psi$ it is natural to expect $m_{(i)} \lesssim \Lambda_{\rm HC}$. 
Note that since all $m_{(i)}$ share the same gap equation and have the same value, the $\SU(5)$ global symmetry is broken down to $\SO(5)$. 
Thus the $\Sp(2N_{\rm HC})$ theory provides a suitable UV completion of the $\SU(5)/\SO(5)$ composite Higgs model.

The $\SU(5)/\SO(5)$ coset contains fourteen Nambu-Goldstone bosons, which under the electroweak group $\SU(2)_L \times \U(1)_Y$ decompose as~\cite{Mrazek:2011iu,Ferretti:2014qta}
\begin{equation}
	{\bf 14} \to {\bf 2}_{\pm \frac{1}{2}} + {\bf 3}_0 + {\bf 3}_{\pm 1}+{\bf 1}_0\,.
	\label{eq:su5so5ngb}
\end{equation}
Besides the Higgs doublet $({\bf 2}_{\pm \frac{1}{2}})$ there are three electroweak triplets and a singlet.
A Higgs potential arises via explicit global symmetry breaking from top quark couplings. In particular, the Higgs field receives negative mass-squared corrections from the top quark Yukawa couplings, leading to electroweak symmetry breaking.
Instead the triplets in~\eqref{eq:su5so5ngb} only receive positive mass-squared loop corrections from gauge bosons, and thus their VEVs are stabilized at zero~\cite{Ferretti:2014qta}. On the other hand, the singlet can receive a mass from an explicit mass term for the $\psi$ hyperfermions.

For the global symmetry breaking pattern of the enlarged color sector, the interaction potential for $\Psi_{\chi}$ is more involved, and will be discussed in Section~\ref{subsec:hc_chi_potential}. Next, we present the case of $\SU(4)_{\rm HC}$, which shares many similarities and provides an alternative hypercolor gauge group. 

\subsubsection{\texorpdfstring{$\SU(4)_{\rm HC}$}{SU(4) hypercolor}}
\label{subsubsec:hc_unitary}

The $\SU(5)/\SO(5)$ composite Higgs model can also arise from the hypercolor gauge group $\SU(4)_{\rm HC}$~\cite{Ferretti:2014qta}. Similar to the $\Sp(2N_{\rm HC})$ theory, this model contains $\psi$ in the antisymmetric representation, which is real, and $\Psi_{\chi}$ in the fundamental representation, which is now complex instead of pseudoreal, as in the symplectic hypercolor case. The hyperfermion content is summarized in Table~\ref{tab:su4_particle_content}.

\begin{table}[H]
    \centering
    \begin{tabular}{|l||c|c|c||c|c|c|}
    \cline{1-7}
                        & $\SU(4)_{\rm HC}$              & $\SU(N+3)$        & $\SU(N)'$     &   $\SU(5)$                & $\U(2N + 3)_L$           & $\U(2N + 3)_R$        \\ \hline 
        $\psi$          & {\bf 6}               & {\bf 1}           & {\bf 1}       &   {\bf 5}                 & {\bf 1}                   & {\bf 1}               \\ \hline
        $\chi$          & {\bf 4}               & {\bf F}           & {\bf 1}       &   \multirow{2}{*}{\bf 1}  & \multirow{2}{*}{\bf F}    & \multirow{2}{*}{\bf 1}   \\
        $\chi'$         & {\bf 4}               & {\bf 1}           & {\bf F}       &                           &                           &                         \\ \cline{1-7}
        $\bchi$         & ${\bf \bar{4}}$       & ${\bf \bar{F}}$   & {\bf 1}       &   \multirow{2}{*}{\bf 1}  & \multirow{2}{*}{\bf 1}    & \multirow{2}{*}{\bf F}  \\
        $\bchi'$        & ${\bf \bar{4}}$       & {\bf 1}           & ${\bf \bar{F}}$ &                         &                           &                         \\ 
    \cline{1-7}
    \end{tabular}
    \caption{The hypercolor fermion representations transforming under the local symmetry, $\SU(4)_{\rm HC}\times \SU(N+3)\times \SU(N)'$ and the global symmetry, $\SU(5)\times \U(2N+3)_L\times \U(2N+3)_R$.}
    \label{tab:su4_particle_content}
\end{table}

The $\SU(4)_{\rm HC}$ gauge group allows for an invariant fermion bilinear
\begin{equation}
    (\psi^a \psi^b) = \epsilon^{ijkl} \psi^a_{ij} \psi^b_{kl} \,,  
    \label{eq:hc_su4_psipsi}
\end{equation}
where $\epsilon^{ijkl}$ is the Levi-Civita tensor and $i, j = 1 \ldots 4$ are $\SU(4)_{\rm HC}$ fundamental indices. To obtain the vacuum structure, a similar four-fermion analysis can be done as in the $\Sp(2N_{\rm HC})$ case. Since the $\SU(5)$ global symmetry of $\psi$ remains the same, and the invariant fermion bilinear \eqref{eq:hc_su4_psipsi} does not affect the global symmetry breaking dynamics, the same $\SU(5)/\SO(5)$ breaking coset follows from Eqs.~\eqref{eq:hc_sp_psi_int}-\eqref{eq:gap}. This coset was also analyzed in Ref.~\cite{Ferretti:2014qta} for $\SU(4)_{\rm HC}$.

However, the enlarged color sector analysis differs from the $\Sp(2N_{\rm HC})$ case since $\Psi_{\chi}$ no longer has a homogeneous representation under the hypercolor gauge group. 
For example, bilinears between hyperfermions $\chi (\bchi)$ and $\chi' (\bchi')$ are not allowed. Furthermore, the global symmetry of $\Psi_{\chi}$ is now restricted to $G_{\SU} = \U(2N + 3)_L \times \U(2N + 3)_R \subset G_{\Sp}$.
Thus, when the enlarged color symmetry is broken, the $\SU(4)_{\rm HC}$ case  is expected to have a different vacuum structure as well as Nambu-Goldstone boson spectrum.

\subsection{Enlarged Color Breaking with Hyperfermion Condensates}
\label{subsec:hc_chi_potential}

\subsubsection{\texorpdfstring{$\SU(N+3)\times \SU(N)'$}{SU(N+3) x SU(N)'} Gauged NJL Model}
\label{subsubsec:color_njl}

Similarly to the $\psi\psi$ condensate, which is responsible for breaking the global $\SU(5)$ symmetry to $\SO(5)$, the colored hyperfermion condensates will be used to break the enlarged color symmetry to $\SU(3)_c\times \SU(N)_D$. 
The hypercolor strong dynamics therefore provides a natural way to break the enlarged color group without the introduction of new elementary scalars in the theory.  

To determine how the enlarged color symmetry is broken we will consider four fermion operators that contain the colored hyperfermions in $\Psi_{\chi}$. 
The most compact way to write down the Lagrangian between these fields is by introducing an auxiliary field similar to $M^{ab}$ in Eq.~\eqref{eq:hc_sp_aux_m}. 
In the case of symplectic hypercolor
\begin{align}
    &\Lag_{\Sp} 
    \supset \frac{1}{2}  M_{\Sp} \Psi^{\dagger}_{\chi} \Psi^{\dagger}_{\chi} + \hc   \notag \\
    &\equiv \frac{1}{2} \left( 
    \chi^{\dagger F} \,\,\, \chi'^{\dagger f} \,\,\, \bchi^{\dagger}_H \,\,\, \bchi'^{\dagger}_h
    \right)    
    \left( \begin{array}{@{}cccc@{}}
        P_{FG}                  & \Phi_{Fg}             & R_F^{\x K}            & \Delta_F^{\x k}    \\
        -(\Phi^T)_{fG}           & P'_{fg}               & \tDelta_f^{\x K}      & {R'}_f^{\x k}        \\
        -(R^T)^H_{\x G}          & -(\tDelta^T)^H_{\x g}  & \tilde{P}^{HK}        & \tPhi^{Hk}   \\
        -(\Delta^T)^h_{\x G}     & -(R'^T)^h_{\x g}       & -(\tPhi^T)^{hK}        & \widetilde{P}'^{hk} 
    \end{array} \right) 
    \left( \begin{array}{c} 
    \chi^{\dagger G} \\ \chi'^{\dagger g} \\ \bchi^{\dagger}_K \\ \bchi'^{\dagger}_k
    \end{array}\right) + \hc \,,
    \label{eq:msp}
\end{align}
where $P,P',\widetilde{P},\widetilde{P}', \Phi,\tPhi, \Delta, \tDelta, R, R'$ are auxiliary fields and uppercase (lowercase) letters denote $\SU(N + 3) \, (\SU(N)')$ fundamental indices. The $P, P', \widetilde{P}$ and $\widetilde{P}'$ fields in the diagonal entries of $M_{\Sp}$ are antisymmetric since $\Psi_{\chi}$ is in the pseudoreal representation. 
The $P, P', \widetilde{P},\widetilde{P}'$ and $\Phi,\tPhi,$ auxiliary fields can be treated as mass mixing terms and can be rotated away using part of the global symmetry $G_{\Sp}$ of $\Psi_{\chi}$. 
After this procedure, the rotated mass matrix is still invariant under $\U(2N + 3)_L \times \U(2N+ 3)_R$, which incidentally is the same as $G_{\SU}$ in the $\SU(4)_{\rm HC}$ hypercolor case. 

In the $\SU(4)_{\rm HC}$ case, the $P$ and $\Phi$ auxiliary fields are automatically forbidden since these fields are not gauge invariant under the new hypercolor. Instead the most general auxiliary field mass matrix becomes complex symmetric
\begin{align}
    M_{\SU} = 
    \left( \begin{array}{@{}cccc@{}}
        0                  & 0             & R            & \Delta    \\
        0           & 0               & \tDelta      & R'        \\
        R^T         & \tDelta^T  & 0        & 0   \\
        \Delta^T     & R'^T       & 0        & 0 
    \end{array} \right)  \,.
\end{align}
So in both cases, the only fields we need to consider are $\Delta, \tDelta$ and $R, R'$. The relevant four-fermion interaction Lagrangian is thus
\begin{align}
    \Lag_{\rm int}  =  \frac{\kappa_R}{2 N_{\rm HC}} ( \chi_F \bchi^G )\Big|_{\tr=0}( \chi^{\dagger F} \bchi^{\dagger}_G )\Big|_{\tr=0}  
                &+ \frac{\kappa_{R'}}{2 N_{\rm HC}} ( \chi'_f  \bchi'^g )\Big|_{\tr=0}( \bchi'^{\dagger f} \bchi'^{\dagger}_g )\Big|_{\tr=0}  \notag \\
                + \frac{\kappa_{\Delta}}{2 N_{\rm HC}} ( \chi_F \bchi'^f  ) ( \chi^{\dagger F} \bchi'^{\dagger}_f ) 
                &+ \frac{\kappa_{\widetilde{\Delta}}}{2 N_{\rm HC}} ( \bchi^F  \chi'_f  ) ( \bchi^{\dagger}_F  \bchi'^{\dagger f} ) \,,
    \label{eq:chi_int_fermion}
\end{align}
where similar to $\kappa_{\psi}$, the four-fermion couplings $\kappa_{R}, \kappa_{R'}, \kappa_{\Delta}, \kappa_{\tDelta}$ are assumed to be real and positive. Note that in the case of $\SU(4)_{\rm HC}$, the $2N_{\rm HC}$ factor in \eqref{eq:chi_int_fermion} and subsequent expressions is replaced with 4. The relevant auxiliary field definitions are
\begin{align}
    {R_F}^G            &= -\frac{\kappa_R}{2N_{\rm HC}}           (\chi_F \bchi^G)\Big|_{\tr = 0}\,,               
    &{R'_f}^g           &= -\frac{\kappa_{R'} }{2N_{\rm HC}}       (\chi'_f  \bchi'^g)\Big|_{\tr = 0}\,,         \notag \\ 
    {\Delta_F}^f        &= -\frac{\kappa_{\Delta}}{2N_{\rm HC}}      (\chi_F \bchi'^f)\,,             \qquad 
    &{\tDelta^F}_{\x f} &= -\frac{\kappa_{\tDelta}}{2N_{\rm HC}}     (\bchi^F   \chi'_f)\,.    
    \label{eq:chi_Delta_R}
\end{align}
The full interaction Lagrangian can now be written as a Yukawa model:
\begin{align}
        \Lag_{\rm int}   = &- \left( R^T ( \chi \bchi ) 
                            + R'^T (\chi' \bchi') 
                            + \Delta^{\dagger} (\chi \bchi')
                            + \tDelta^{\dagger} (\bchi \chi') + \hc
                            \right) \notag  \\
                           &- \frac{2 N_{\rm HC}}{ \kappa_{R} } \tr (R^2) 
                            - \frac{2 N_{\rm HC}}{ \kappa_{R'} } \tr (R'^2)
                            - \frac{2 N_{\rm HC}}{ \kappa_{\Delta} } \tr (\Delta \Delta^{\dagger})   
                            - \frac{2 N_{\rm HC}}{ \kappa_{\tDelta} } \tr (\tDelta \tDelta^{\dagger}) \,.
        \label{eq:chi_yukawa}
\end{align}
Note that the traces $\chi_F \bchi^F$ and $\chi'_f \bchi^f$ have not been included in Eqs.~\eqref{eq:chi_int_fermion}-\eqref{eq:chi_yukawa}. These fields are singlets under $\SU(N + 3)$ and $\SU(N)'$, and thus will be omitted from the $\Psi_{\chi}$ effective potential because they do not break any gauge symmetries.
Nonetheless, these traces can still acquire VEVs through non-perturbative effects or can have arbitrary bare mass terms $m_{\chi} \chi_F \bchi^F$ and $m_{\chi'} \chi'_f \bchi'^f$, that break the axial symmetry $\U(1)^A_{\chi}$ and $\U(1)^A_{\chi'}$ of $\SU(N+3)$ and $\SU(N)'$ respectively (see Appendix~\ref{app:B}). If these mass terms are set to zero, then these symmetries are still anomalous due to the enlarged color instanton contributions at the scale $\Lambda_{\rm HC}$. We denote the Nambu-Goldstone bosons associated with the breaking of these symmetries $\sigma$ and $\sigma'$, respectively.
In our model, we will assume that $m_{\chi, \chi'}$ are nonzero, but small compared to $\Lambda_{\rm HC}$. 
For simplicity, we also assume that $m_{\chi, \chi'}$ are real, so that the topological aspects of the hypercolor and the enlarged color sectors are entirely separated. The $\sigma$ and $\sigma'$ masses obtain contributions, either from these explicit breaking mass terms, or from hypercolor or enlarged color instantons. Therefore, it is reasonable to assume that these singlets will be as heavy as the $\Lambda_{\rm HC}$ scale.

\subsubsection{Effective Potential for the Auxiliary Fields}

The next step is to obtain the effective potential by integrating out the fermions. First we derive the gap equation for the auxiliary field, $\Delta$. To a good approximation (see appendix~\ref{app:A})
\begin{align}
    V_{\text{eff}} \approx& \,\frac{2N_{\rm HC}}{\kappa_{\Delta}} \tr (\Delta \Delta^{\dagger})  \notag \\
        &- \frac{N_{\rm HC}}{8 \pi^2} \tr \left(
            \Lambda_{\rm UV}^2  \Delta \Delta^{\dagger} - ( \Delta \Delta^{\dagger} )^2 \log \left(1+\frac{ \Lambda_{\rm UV}^2}{ \Delta \Delta^{\dagger} }  \right)
        + \Lambda_{\rm UV}^4 \log \left(1+\frac{ \Delta \Delta^{\dagger} }{\Lambda_{\rm UV}^2}  \right)\right) \notag \\
        &- \frac{N_{\rm HC}}{16\pi^2} \tr
        \left( \left( - \frac{2 \Lambda_{\rm UV}^2 }{\Lambda_{\rm UV}^2 + \Delta \Delta^{\dagger}} + 
        \log \left(1+\frac{ \Lambda_{\rm UV}^2} { \Delta \Delta^{\dagger} }
         \right)\right) \Delta R' \tDelta R 
        + \hc \right) \,,
    \label{eq:delta_veff}
\end{align}
where the first two terms are the standard Coleman-Weinberg potential for $\Delta$. The last term is novel, arising from loops that contain $\Delta$ as well as other fields $R,R'$. 
In this approximation, other auxiliary fields have been kept only to first order. 
Using the global symmetry $ \SU(N + 3)_{\chi} \times \SU(N + 3)_{\bchi} \times \SU(N)_{\chi'} \times \SU(N)_{\bchi'} \subset G_{\SU}$, one can simultaneously decompose either the pair $\Delta$, $\tDelta$ or the pair $R$, $R'$, but not both pairs of fields. Without loss of generality, we choose to apply the singular value decomposition on $\Delta$, $\tDelta$, which gives
\begin{equation}
    \Delta = U_{\chi} \Delta_D V^{\dagger}_{\bchi'},  \qquad 
    \tDelta = U_{\bchi} \tDelta_{\widetilde{D}} V^{\dagger}_{\chi'} \,. 
\end{equation}
The non-negative singular values are correspondingly $\Delta_{n}$ and $\tDelta_{n}$ ($n=1\dots N$) 
\begin{equation}
    \Delta_D = 
    \left( \begin{array}{@{}ccc@{}} 
   \, & \, & \, \\
    \, & \text{diag} \left( \Delta_{1}, \ldots, \Delta_{N} \right) & \, \\
    \, & \, & \, \\ 
    \cmidrule{1-3}
    \, & {\bf 0}_{3\times N} & \, 
    \end{array} \right) \,, \qquad
    \tDelta_D = 
    \left( \begin{array}{@{}ccc@{}} 
   \, & \, & \, \\
    \, & \text{diag} \left( \tDelta_{1}, \ldots, \tDelta_{N} \right) & \, \\
    \, & \, & \, \\ 
    \cmidrule{1-3}
    \, & {\bf 0}_{3\times N} & \, 
    \end{array} \right) 
    \label{eq:Ddiagonal}
\end{equation}
For the eigenvalues, $\Delta_{n}$, the mass gap equation is now coupled with other fields
\begin{equation}
    \frac{1}{\zeta_{\Delta}} \equiv \frac{4\pi^2}{ \kappa_{\Delta} \Lambda_{\rm UV}^2 } =   
       1 - \frac{2 \phi^{\Delta}_{n} }{\Delta_{n}^2 ( \Lambda_{\rm UV}^2 + \Delta_{n}^2) }
    - \frac{\Delta_{n}^2}{\Lambda_{\rm UV}^2}\left( 1 - \frac{ \phi^{\Delta}_{n} }{ \Delta_{n}^4 } \right)  \log \left(1+\frac{\Lambda_{\rm UV}^2} {\Delta_{n}^2}\right)\,, 
    \label{eq:gap_modified}
\end{equation}
where $\phi^{\Delta}$ controls the amount of mixing with other fields. By setting $\phi^{\Delta} = 0$ one recovers the original gap equation in~\eqref{eq:gap}. The gap equations for $\Delta, \tDelta, R$ and $R'$ only differ by the definition of $\phi$ 
\begin{align}
    (\phi^{\Delta})_{n} &= \Delta_{n} \tDelta_{n} \sum_{m = 1}^N R^{\dagger n}_{\x \x m} R'^{\dagger m}_{\x \x n} \,, \quad& 
    (\phi^{\tDelta})_{n} &= \Delta_{n} \tDelta_{n} \sum_{m = 1}^N R'^{\dagger n}_{\x \x m} R^{\dagger m}_{\x \x n} \,, \notag \\
    (\phi^{R})_{n}^{\x m} &= R_{n}^{\x m} \Delta_{m} {R'}_{m}^{\x n} {\tDelta}_{n} \,, \quad & 
    (\phi^{R'})_{n}^{\x m} &= R'^{\x m}_{n} \tDelta_{m} R_{m}^{\x n} \Delta_{n}  \,.
    \label{eq:gap_mixing}
\end{align}
The $R_F^{\x G}$ components where $F,G > N$ have not been included, since the mixing term $\phi^{R}$ for these components is zero, i.e. these components do not mix with other fields, thus satisfy the original gap equation. 

For simplicity we assume $\kappa_{\Delta} = \kappa_{\tDelta}$ and $\kappa_{R} = \kappa_{R'}$, such that there are only two parameters controlling the symmetry breaking pattern. In this case, one can numerically solve the system of equations in~\eqref{eq:gap_modified} by further assuming that $R_n^{\x m}$ have the same value for $n \ne m$. 

    \begin{figure}[H]
    	\centering
    	\begin{subfigure}[t]{0.6\textwidth}
    	\includegraphics[width=\textwidth]{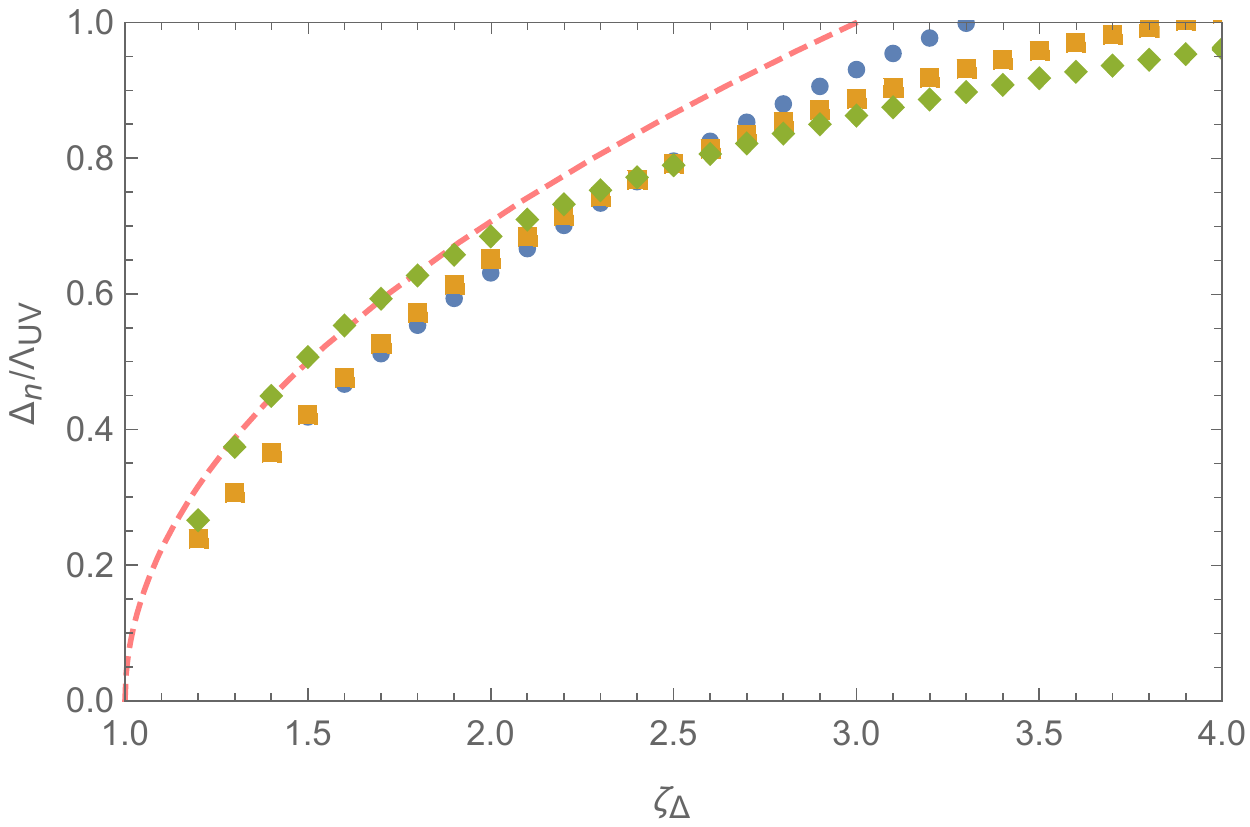}
     	    \subcaption{}
     	    \label{fig:m_delta}
     	\end{subfigure}
        ~
     	\begin{subfigure}[t]{0.62\textwidth}
        \includegraphics[width=\textwidth]{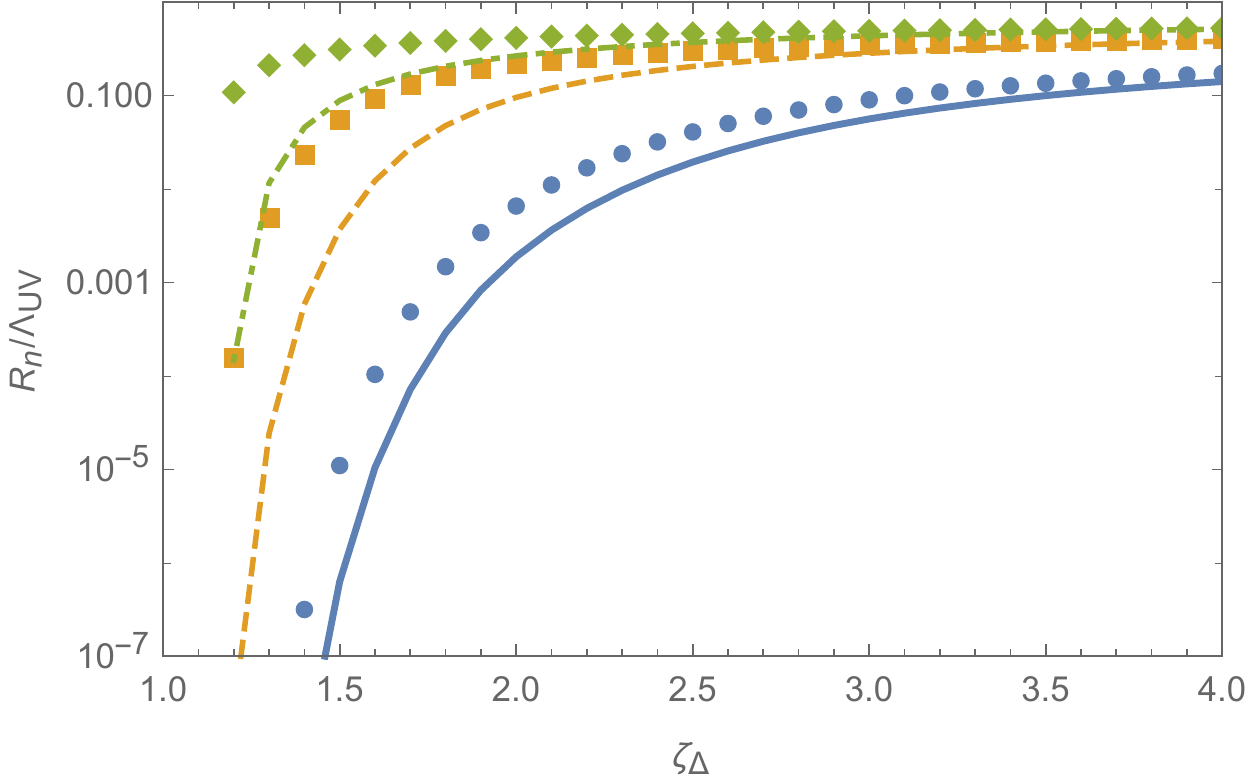}
            \subcaption{}
            \label{fig:m_phi}
     	\end{subfigure}
     	\caption{Numerical values for (a) $\Delta_{n}$ and (b) $R_{n}^{\x m}$ as functions of $\zeta_{\Delta}$ in the regime $\zeta_{\Delta} > 1, \zeta_R < 1$. The blue dots, yellow squares and green diamonds correspond to $\zeta_{R} = 0.2, 0.5,$ and $0.8$, respectively. In (a), the dashed red curve represents the analytic approximation for $\Delta_{n}$ without mixing terms given in Eq.~\eqref{eq:delta_analytic}. In (b), the blue (solid), yellow (dashed) and green (dot-dashed) curves represent the analytic approximation for $R_n^{\x m}$ given in Eq.~\eqref{eq:R_analytic}. This approximation works better for small $\zeta_R$ (lower curves). }
     	\label{fig:num_delta}
    \end{figure}

\begin{figure}[H]
    \centering
 	\includegraphics[width = 0.5 \textwidth]{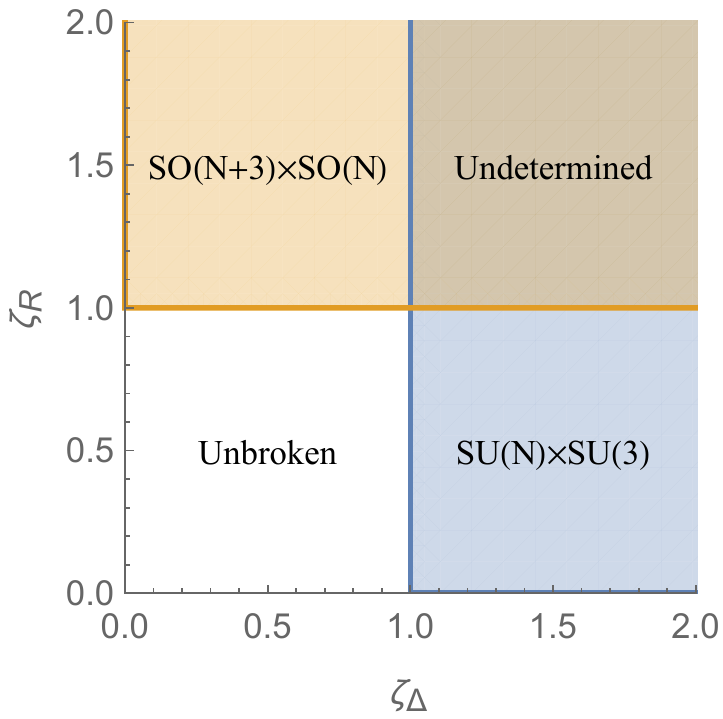}
 	\caption{The enlarged color phase diagram, where $\zeta_{\Delta} = \kappa_{\Delta} \Lambda_{\rm UV}^2/(4 \pi^2)$ and $\zeta_{R} = \kappa_{R} \Lambda_{\rm UV}^2/(4 \pi^2)$ are dimensionless four-fermion couplings of $\Delta$ and $R$ respectively.}
    \label{fig:phase}
\end{figure}

When $\zeta_{\Delta}$ and $\zeta_{R}$ are less than one, both $\Delta$ and $R$ stabilize at zero. Thus in this regime, the enlarged color symmetry is unbroken. 
In the regime where $\zeta_{\Delta} > 1$ and $\zeta_{R} < 1$, the numerical result (see Figure~\ref{fig:num_delta}) indicates that the mixing term $\phi^{\Delta}$ has little effect on the mass gap equation for $\Delta$. Thus for all $n = 1 \ldots N$, the eigenvalues $\Delta_{n}$ share the same analytic expression, derived from the original gap equation,
\begin{equation}
    \frac{ \Delta_{n} }{ \Lambda_{\rm UV} } \approx \sqrt{ \frac{\zeta_{\Delta} - 1}{2} }  \,.
    \label{eq:delta_analytic}
\end{equation}
Without the mixing term, $R_n^{\x m}$ should stabilize at zero given that $\zeta_{R} < 1$. With the new $\phi^R$ term, $R_n^{\x m}$ is now allowed to be nonzero, though turned out to be exponentially suppressed. The analytic approximation for $R$ derived from~\eqref{eq:gap_modified} is
\begin{equation}
    R_{n}^{\x m} \approx \Lambda_{\rm UV} \exp \left( -  \frac{ \Lambda_{\rm UV}^2 }{ 2 \zeta_{R}  \Delta_{n} \tDelta_{m} } \right).
    \label{eq:R_analytic}
\end{equation}
Thus in this parameter subspace, $\Delta$ is the relevant VEV. 
Since $\Delta$ is a bifundamental under $\SU(N + 3) \times \SU(N)'$, the enlarged color is broken to the diagonal subgroup $\SU(N)_D \times \SU(3)_c$. 
When $\zeta_{R} > 1$ and $\zeta_{\Delta} < 1$, it is straightforward to see that the field $\Delta$ and $R$ now have reverse roles. The auxiliary field $R (R')$, which is an adjoint under $\SU(N + 3) (\SU(N)')$, is now the relevant VEV, and thus breaking the symmetry to $\SO(N + 3) \times \SO(N)$. 
The phase diagram summarizing this result is shown in Figure~\ref{fig:phase}. 

When $\zeta_{\Delta}$ and $\zeta_{R}$ are both greater than one, both auxiliary fields $\Delta, R$ can have nontrivial VEVs of order $\Lambda_{\rm UV}$. In this case the approximation $\zeta_{\Delta} \approx \zeta_{\tDelta}$ and $\zeta_R \approx \zeta_{R'}$ is too crude to capture the symmetry breaking pattern of the model. It requires a more in-depth analysis beyond the scope of this paper, and thus this parameter regime is labelled as undetermined in Figure~\ref{fig:phase}. 
Nevertheless for the rest of this paper, we will assume $\zeta_{\Delta, \tDelta} > 1$, while $\zeta_{R, R'} < 1$, such that $\Delta$ and $\tDelta$ are the only relevant VEVs. In the next section we show in detail how these VEVs imply the enlarged color breaking $\SU(N + 3) \times \SU(N) \to \SU(N)_D \times \SU(3)_c$.

\subsection{Enlarged Color Symmetry Breaking}
\label{subsec:Enlargedcolorbreaking}

To show that $\Delta_{F}^{\x f}$, which transforms in the fundamental (antifundamental) representation of $\SU(N + 3)$~$(\SU(N)')$, provides the desired local symmetry breaking pattern, we next consider the $\Delta$ kinetic term. This term is a proxy for the hyperfermion dynamics, which produces dynamical Nambu-Goldstone bosons that will be eaten by the spontaneously broken gauge bosons.  The covariant derivative with respect to the enlarged color group is
\begin{equation}
    \label{eq:covariant_der}
    D_{\mu} \Delta = \partial_{\mu} \Delta  - i T^A G_{\mu}^A \Delta
    + i  \Delta T'^{a}  G'^a_{\mu}\,,
\end{equation}
where $A, a$ are $\SU(N + 3), \SU(N)'$ indices, $G^A, G'^a$ are the corresponding gluons, and $T^A, T'^a$ are the corresponding group generators, which are $(N + 3) \times (N + 3)$, $N\times N$ traceless, Hermitian matrices, respectively. The $\SU(N + 3)$ generators can be grouped as $T^A= \{T_N^a, T^b, T_3^c, T_1 \}$ where
\begin{align}
    \{ T_N^a \} = 
    \left\{ \left( \begin{array}{@{}c|c@{}}
    \begin{matrix}
      \, & \, & \, \\
      \, & \SU(N) & \, \\
      \, & \, & \, 
    \end{matrix} 
      & {\bf 0}_{N\times 3} \\
    \cmidrule[0.4pt]{1-2}
    {\bf 0}_{3\times N} & {\bf 0}_{3} \\
    \end{array} \right) \right\}, \,\,\,\, 
    &\{ T^b \} = \frac{1}{2} \left\{
    \left( \begin{array}{@{}c|c@{}}
    \begin{matrix}
      \, & \, & \, \\
      \, & {\bf 0}_{N} & \, \\
      \, & \, & \, 
    \end{matrix} 
      & [1]\left([i]\right) \\
    \cmidrule[0.4pt]{1-2}
    [1]([-i]) & {\bf 0}_{3} \\
    \end{array} \right) \right\}, \notag \\
    \{ T_3^c \} = 
     \left\{ \left( \begin{array}{@{}c|c@{}}
    \begin{matrix}
      \, & \, & \, \\
      \, & {\bf 0}_{N} & \, \\
      \, & \, & \, 
    \end{matrix} 
      & {\bf 0}_{N\times 3} \\
    \cmidrule[0.4pt]{1-2}
    {\bf 0}_{3\times N} & \SU(3) \\
    \end{array} \right) \right\}, \,\,\,\,
    &T_1 = \frac{1}{\sqrt{6N(N+3)}}  \left( \begin{array}{@{}c|c@{}}
    \begin{matrix}
      \, & \, & \, \\
      \, & 3 \, \id_N & \, \\
      \, & \, & \, 
    \end{matrix} 
      & {\bf 0}_{N\times 3} \\
    \cmidrule[0.4pt]{1-2}
    {\bf 0}_{3\times N} & - N \, \id_3 \\
    \end{array} \right),
    \label{eq:TAgroup}
\end{align}
with the normalization $\tr T^A T^B = \frac{1}{2} \delta_{AB}$. Thus, $\{T_N^a\}$ consists of $N^2 - 1$ generators containing the subgroup $\SU(N)$;  $\{T^b\}$ consists of $6N$ sparse matrices which have a single $1$ or $i$ in any of the $3N$ positions in the upper right region, and a 1 or $-i$ in the corresponding symmetric position of the lower left region; $\{ T_3^c \}$ consists of eight generators containing the subgroup $\SU(3)$; and $T_1$ is the generator of the \U(1) subgroup that commutes with all other elements of $\{T_N^a\}$ and $\{T_3^c\}$. 
Similarly, the $\SU(N + 3)$ gauge bosons  can also be grouped according to the generators in \eqref{eq:TAgroup}, namely $G_\mu^A =\{ G_{N\mu}^a, G_\mu^b, G_{3\mu}^c, G_{1\mu}\}$. Assuming that $\Delta_n \simeq f_{\rm HC} = \Lambda_{\rm HC}/(4 \pi)$, when the hypercolor group confines, $\Delta$ obtains a VEV  
\begin{equation}
    \braket{\Delta}_F^{\x f} = 
    f_{\rm HC} 
    \left( \begin{array}{@{}ccc@{}} 
   \, & \, & \, \\
    \, & \id_N & \, \\
    \, & \, & \, \\ 
    \cmidrule{1-3}
    \, & {\bf 0}_{3\times N} & \, 
    \end{array} \right) \,, 
    \label{eq:delta_vev}
\end{equation}
which generates mass terms for the spontaneously broken gauge bosons: 
\begin{align}
    \tr D_{\mu} \Delta D^{\mu} \Delta^{\dagger} =
        \frac{1}{2} f_{\rm HC}^2 (G_{N\mu}^a - G'^a_{\mu})^2 + 
        \frac{1}{2} f_{\rm HC}^2 (G_{\mu}^b)^2 + 
        \frac{1}{2} \frac{ 3f_{\rm HC}^2 }{ (N+3) } (G_{1\mu})^2\,.
    \label{eq:massgivegbs}
\end{align}
The massive eigenstates are $\frac{1}{ \sqrt{2} } ( G_{N\mu}^a - G'^a_{\mu} ), G_{\mu}^b$ and $G_{1\mu}$, while the massless gauge bosons are: \begin{equation}
    G^{a}_{D\mu} = \frac{1}{\sqrt{2}}(G^a_{N\mu} + G'^a_{\mu})\,, \qquad
    G^c_{3\mu}\,, 
    \label{eq:masslessgbs}
\end{equation}
which are respectively associated with the gauge groups $\SU(N)_D$ and $\SU(3)_c$ where $a=1\dots N^2-1$ and $c=1\dots 8$. The mass of the $\U(1)$ gauge boson, $G_{1\mu}$ is of order $1/N$, which vanishes in the large $N$ limit. 

Since a part of the enlarged color global symmetry is weakly gauged, there are $N(N+6)$ Nambu-Goldstone bosons from the spontaneous symmetry breaking, which are eaten by the massive gauge bosons, $\frac{1}{ \sqrt{2} } ( G_{N\mu}^a - G'^a_{\mu} ), G_{\mu}^b$, and  $G_{1\mu}$ (see Appendix~\ref{app:B}). This is unlike the spontaneous breaking of the $\SU(5)$ global symmetry, where the Higgs doublets remain as Nambu-Goldstone bosons. In other words, the hypercolor dynamics breaks the enlarged color group, but does not break the Standard Model electroweak gauge group. 

We can now obtain the $\SU(N)_D$ and $\SU(3)_c$ coupling constants at the scale $\Lambda_{\rm HC}$ by considering the gauge kinetic term 
\begin{equation}
    \frac{1}{4 g_{N+3}^2} F^{A\mu\nu} F^A_{\mu\nu} + 
    \frac{1}{4 g'^2} F'^{a\mu\nu} F'^a_{\mu\nu} = 
    \frac{1}{4} \left( \frac{1}{g_{N+3}^2} + \frac{1}{g'^2} \right) F_D^{a\mu\nu } F^a_{D\mu\nu} + \frac{1}{4g_{N+3}^2} F_3^{c\,\mu\nu} F^c_{3\mu\nu} \,,
    \label{eq:gauge_split}
\end{equation}
where in the $F_D$ term, $G_{D\mu}^a$ has been rescaled by $\sqrt{2}$.
This leads to the identification
\begin{equation}
     \frac{1}{\alpha_D(\Lambda_{\rm HC})}=\frac{1}{\alpha_{N+3}(\Lambda_{\rm HC})} + \frac{1}{\alpha'(\Lambda_{\rm HC})}\,,
     \label{eq:alphaprime}
\end{equation}
and $\alpha_c(\Lambda_{\rm HC}) = \alpha_{N+3}(\Lambda_{\rm HC})$ where $\alpha_i= g_i^2/4\pi$. This breaking generalizes previous results for $\SU(3)\times \SU(3)$~\cite{ArkaniHamed:2001ca,Cheng:2001nh,Agrawal:2017ksf}. Similarly, at the scale $\Lambda_{\rm HC}$ the corresponding $\theta$ terms of the gauge fields can be identified as
\begin{equation}
   \theta_{N+3} F^{A\mu\nu} \widetilde{F}^A_{\mu\nu} +  \theta' F'^{a\mu\nu} \widetilde{F}'^a_{\mu\nu} =
     (\theta_{N+3} + \theta') F_D^{a\mu\nu } \widetilde{F}^a_{D\mu\nu} + 
    \theta_{N+3} F_3^{c\,\mu\nu} \widetilde{F}^c_{3\mu\nu} \,,
    \label{eq:theta_lag}
\end{equation}
where $\widetilde{F}_{\mu\nu}^i = \frac{1}{2} \epsilon_{\mu\nu\rho\sigma} F^{i\rho\sigma}$ $(i=A,a,c)$ and $\epsilon_{0123} = \epsilon^{0123} = + 1$. The new gauge fields of $\SU(N)_D$ and $\SU(3)_c$ then inherit $\theta$ terms:
\begin{equation}
    \theta_D = \theta_{N+3} + \theta'\,, \qquad \qquad \theta_c = \theta_{N+3}\,.
    \label{eq:theta}
\end{equation}
There are also phases in the fermion sector that will contribute to an effective $\theta$-term. We next discuss how the Standard Model quarks are embedded into the enlarged color sector.

\subsection{Embedding the Standard Model Quarks}
\label{subsec:SMquarkembedding}

The Standard model quarks, $q,{\bar u},{\bar d}$ are embedded into the larger representations $Q, {\bar U}, {\bar D}$ of the enlarged color group as shown in Table~\ref{tab:quark_content}.
After $\SU(N+3)\times \SU(N)'$ is spontaneously broken, the Standard Model quarks remain massless, while the $\SU(N)_D$ components, ${\bf q}, {\bf {\bar u}}, {\bf {\bar d}}$ pair up with ${\bf \bar{q}'},{\bf u'},{\bf d'}$ to obtain vector-like masses of order $f_{\rm HC}$. For example, consider the gauge interactions of the quark parent $Q$ and $\bar{q}'$
\begin{align}
    Q^\dagger & \bar{\sigma}_{\mu} \left( i\partial^{\mu} + G^{A \mu} T^A \right) Q +
     \bar{q}'^\dagger \bar{\sigma}_{\mu} \left( i\partial^{\mu} - G'^{a \mu} T'^a \right)  \bar{q}'  \notag \\ 
    \rightarrow  &\,i {\bf q}^\dagger \bar{\sigma}_{\mu} \partial^\mu {\bf q} + i\bar{q}'^\dagger \bar{\sigma}_{\mu} \partial^\mu \bar{q}' + 
        G_D^{a \mu} \left( {\bf q^\dagger} \bar{\sigma}_{\mu} T_D^a {\bf q} - \bar{q}'^\dagger \bar{\sigma}_{\mu} T_D^a \bar{q}' \right) + i q^\dagger \bar{\sigma}_{\mu} \partial^\mu q + G^{c \mu}_3 q^\dagger \bar{\sigma}_{\mu} T^c_3 q\,,
	\label{eq:fermion_gauge}
\end{align}
where $T_D^a$ is an $\SU(N)_D$ generator, $A$, $a$, and $c$ are $\SU(N+3)$, $\SU(N)'$, and $\SU(3)_c$ indices respectively, and $G_D^{a\mu}$ has been rescaled by $\sqrt{2}$ as in \eqref{eq:gauge_split}. In addition, there is a mass term generated by an effective $\Delta$ coupling:
\begin{equation}
\label{eq:veccoupling}
    y_Q \Delta_F^{\dagger f} Q^F \bar{q}'_{f} + \hc \rightarrow y_Q f_{\rm HC} \, {\bf q}^f {\bf \bar{q}'}_f + \hc \,,
\end{equation}
where $y_Q$ is a $3\times 3$ Yukawa coupling matrix in family space with family indices omitted. 
There are also similar couplings for the right-handed quarks ${\bar U}$ and ${\bar D}$, with respective Yukawa coupling matrices $y_U$ and $y_D$. 

A similar decomposition occurs for the hyperfermions $\chi$, since they are charged under the enlarged color group (see Table~\ref{tab:sp_particle_content} or \ref{tab:su4_particle_content}). Let $\chi = (\bm{\chi}, \chi_c)$ and $\bchi = (\bm{\bchi}, \bchi_c)$, where the first (second) component corresponds to the subgroup generated by $T^a_N$ ($T_3^c$) (see Eq.~\eqref{eq:TAgroup}). Given that the pairs $\bm{\chi}, \bchi'$ and $\bm{\bchi}, \chi'$ form the condensate \eqref{eq:delta_vev}, we obtain the Dirac mass terms
\begin{equation}
    (\Delta^{\dagger})^{\x F}_{f} \chi_F \bchi'^f + (\tDelta^{\dagger})_{\x F}^{f} \bchi^F \chi'_f + \hc = f_{\rm HC} \bm{\chi}_f \bchi'^f + f_{\rm HC} \bm{\bchi}^f \chi'_f + \hc \,,
\end{equation}
while the $\SU(3)_c$ states $\chi_c, \bchi_c$ remain massless. These colored hyperfermions eventually become part of cubic invariants which are identified with the top partners at low energy~\cite{Ferretti:2013kya}. 
The top partners are part of the partial compositeness mechanism for generating the top quark mass. 

More generally, partial compositeness near the hyperfermion confinement scale is responsible for generating the Standard Model flavor structure, although the specific details are model dependent. In order to simplify the analysis relevant for the strong CP problem, we will assume that the global symmetry associated with the electroweak sector is broken by the hyperfermion dynamics at a higher scale compared to the color breaking scale\footnote{In fact this was argued to occur in the $\SU(4)_{\rm HC}$ hypercolor model of Ref.~\cite{Ferretti:2014qta}.}. This means that there is an effective Yukawa coupling of the Higgs to the quark parents. After color breaking, this Yukawa coupling becomes 
\begin{align}
    y_u^{\rm SM} H Q {\bar U} &+ y_d^{\rm SM} H^\dagger Q {\bar D} + \hc \notag \\
    &\to \left( y_u^{\rm SM} H q {\bar u} + y_d^{\rm SM} H^\dagger q {\bar d} + \hc \right) + 
    \left( y_u^{\rm SM} H {\bf q} {\bf {\bar u}} + y_d^{\rm SM} H^\dagger {\bf q} {\bf {\bar d}} + \hc \right)
    \,,
    \label{eq:yukawacoupling}
\end{align}
where $H$ is an $\SU(2)_L$ doublet with $Y=\frac{1}{2}$, and $y_{u,d}^{\rm SM}$ are the Standard Model $3\times 3$ Yukawa coupling matrices with family indices omitted.
Since $q$, $\bar{u}$ and $\bar{d}$ do not receive any mass contribution from $\Delta$ in Eq.~\eqref{eq:veccoupling}, their masses are solely obtained from the Higgs VEV. On the other hand, ${\bf q}, {\bf {\bar u}}$ and ${\bf {\bar d}}$ already obtain Dirac masses of order $f_{\rm HC}$, which are much larger than the contribution they receive from the Higgs Yukawa coupling.

\begin{table}[H]
\centering
\begin{tabular}{c|cccc|cc||cc}
\cline{1-3} \cline{5-9}
                         & $\SU(N + 3)$                                     & $\SU(N)'$                       &  &                     & $\SU(3)_c$                      & $\SU(N)_D$                      & $\SU(2)_L$             & $\U(1)_Y$        \\ \cline{1-3} \cline{5-9} 
\multirow{2}{*}{$Q$}   & \multirow{2}{*}{{\tiny$\yng(1)$}}          & \multirow{2}{*}{${\bf 1}$}     &  & $q$               & {\tiny$\yng(1)$}          & ${\bf 1}$                      & {\tiny$\yng(1)$} & $\frac{1}{6}$   \\
                         &                                                 &                                &  & ${\bf q}$ & ${\bf 1}$                      & {\tiny$\yng(1)$}          & {\tiny$\yng(1)$} & $\frac{1}{6}$   \\ \cline{1-3} \cline{5-9} 
\multirow{2}{*}{${\bar U}$} & \multirow{2}{*}{{\tiny$\overline{\yng(1)}$}} & \multirow{2}{*}{${\bf 1}$}     &  & ${\bar u}$             & {\tiny$\overline{\yng(1)}$} & ${\bf 1}$                      & ${\bf 1}$             & $ - \frac{2}{3}\,\,\,\,$ \\
                         &                                                 &                                &  & ${\bf {\bar u}}$ & ${\bf 1}$                      & {\tiny$\overline{\yng(1)}$} & ${\bf 1}$             & $ - \frac{2}{3}\,\,\,\,$ \\ \cline{1-3} \cline{5-9} 
\multirow{2}{*}{${\bar D}$} & \multirow{2}{*}{{\tiny$\overline{\yng(1)}$}} & \multirow{2}{*}{${\bf 1}$}     &  & ${\bar d}$             & {\tiny$\overline{\yng(1)}$} & ${\bf 1}$                      & ${\bf 1}$             & $\frac{1}{3}$   \\
                         &                                                 &                                &  & ${\bf {\bar d}}$ & ${\bf 1}$                      & {\tiny$\overline{\yng(1)}$} & ${\bf 1}$             & $\frac{1}{3}$   \\ \cline{1-3} \cline{5-9} 
$\bar{q}'$                   & ${\bf 1}$                                       & {\tiny$\overline{\yng(1)}$} &  & ${\bf \bar{q}'}$        & ${\bf 1}$                      & {\tiny$\overline{\yng(1)}$} & {\tiny$\yng(1)$} & $ - \frac{1}{6}\,\,\,\,$ \\ \cline{1-3} \cline{5-9} 
$u'$                 & ${\bf 1}$                                       & {\tiny$\yng(1)$}          &  & ${\bf u'}$        & ${\bf 1}$                      & {\tiny$\yng(1)$}          & ${\bf 1}$             & $\frac{2}{3}$   \\ \cline{1-3} \cline{5-9} 
$d'$                 & ${\bf 1}$                                       & {\tiny$\yng(1)$}          &  & ${\bf d'}$        & ${\bf 1}$                      & {\tiny$\yng(1)$}          & ${\bf 1}$             & $ - \frac{1}{3}\,\,\,\,$ \\ \cline{1-3} \cline{5-9} 
\multirow{5}{*}{$\xi$} & \multirow{5}{*}{\bf Ad} & \multirow{5}{*}{${\bf 1}$}        &  & $\psi'_c$                               & {\tiny$\yng(1)$} & {\tiny$\overline{\yng(1)}$} & ${\bf 1}$ & 0 \\ 
                        &                                                    &                                   &  & $\bpsi'_c$                               & {\tiny$\overline{\yng(1)}$} & {\tiny$\yng(1)$} & ${\bf 1}$ & 0 \\ 
                        &                                                    &                                   &  & $\lambda_c$                              & {\bf Ad} & ${\bf 1}$                         & ${\bf 1}$ & 0 \\
                        &                                                    &                                   &  & $\lambda_D$                              & ${\bf 1}$                         & {\bf Ad} & ${\bf 1}$ & 0 \\ 
                        &                                                    &                                   &  & $\nu'$                              & ${\bf 1}$ & ${\bf 1}$                         & ${\bf 1}$ & 0 \\                        \cline{1-3} \cline{5-9} 
$\psi'$                 & ${\bf 1}$                                          & {\tiny$\yng(1)$} &  & $\psi'$         & ${\bf 1}$                         & {\tiny$\yng(1)$} & ${\bf 1}$ & 0 \\ \cline{1-3} \cline{5-9} 
$\bpsi'$         & ${\bf 1}$                                          & {\tiny$\overline{\yng(1)}$} &  &  $\bpsi'$ & ${\bf 1}$                         & {\tiny$\overline{\yng(1)}$} & ${\bf 1}$ & 0 \\ \cline{1-3} \cline{5-9} 
\end{tabular}
\caption{The Weyl fermion content of the model, including the Standard Model quarks ($q,{\bar u},{\bar d}$), charged under the various gauge groups. All listed fermions are singlets of either hypercolor gauge groups. The Weyl fermion, $\xi$ is listed as an adjoint of $\SU(N+3)$, but as discussed in the text, can also be in a pseudoreal representation of $\SU(N+3)$ for $N=3,7,\dots$.}
\label{tab:quark_content}
\end{table}

Since the $\SU(N)'$ quarks, $\bar{q}',u', d'$ have conjugate Standard Model quantum numbers they can also couple to the Standard Model Higgs field, above the color breaking scale.
The effective Higgs couplings are:
\begin{equation}
\label{eq:Yukawaprime}
    \Delta \Lag' = y_u' \widetilde{H}^\dagger \bar{q}' u' + y_d' \widetilde{H} \bar{q}' d' + \hc
    \rightarrow y_u' \widetilde{H}^\dagger {\bf \bar{q}'} {\bf u'} + y_d' \widetilde{H} {\bf \bar{q}'} {\bf d'} + \hc
    \,,
\end{equation}
where ${\widetilde H}\equiv i\sigma_2 H$ and $y_{u,d}'$ are $3\times 3$ Yukawa coupling matrices in family space with family indices omitted. When the Higgs field obtains a VEV, the Yukawa masses generated in \eqref{eq:Yukawaprime} are negligible compared to the vector-like masses obtained from \eqref{eq:veccoupling}.
Since the Higgs is composite and the primed fermions are elementary, the Yukawa couplings are naturally suppressed. Later they will play an important role in the small instanton contribution to the axion mass. 

Furthermore, after enlarged color and electroweak symmetry is spontaneously broken, the fermion mass term can be organized as follows
\begin{equation}
    \Lag_{\rm Yukawa} = 
    \frac{v}{\sqrt{2}} \left( \begin{array}{cc} 
    {\bar u} &  {\bar d}
    \end{array}\right) 
    Y^{\rm SM} 
    \left( \begin{array}{c} 
    u \\ d 
    \end{array}\right) + 
    f_{\rm HC}
    \left( \begin{array}{cccc} 
    {\bf \bar{u}'} & {\bf {\bar u}} & {\bf \bar{d}'} & {\bf {\bar d}}
    \end{array}\right)
    Y^D
    \left( \begin{array}{c} 
    {\bf u} \\ {\bf u'} \\ {\bf d} \\ {\bf d'} 
    \end{array} \right)   + \hc \,,
    \label{eq:low_energy_Yukawa}
\end{equation}
where $\braket{H}=\frac{1}{\sqrt{2}}\icol{0\\v}$. For convenience, we have written out the $\SU(2)_L$ doublet components as $q = \icol{u\\d}$, ${\bf q} = \icol{{\bf u}\\{\bf d}}$ and ${\bf \bar{q}'} =\icol{{\bf \bar{u}'}\\{\bf \bar{d}'}}$.
The Yukawa coupling matrices are given by
\begin{equation}
    Y^{\rm SM} = 
    \left( \begin{array}{@{}cc@{}}
        y_u^{SM} & 0 \\
        0 & y_d^{SM} 
    \end{array} \right)  \,,
    \qquad 
    Y^D = 
    \left( \begin{array}{@{}cccc@{}}
        y_Q & 0 &  0 & \frac{v}{\sqrt{2}f_{\rm HC}} \, y'_d  \\
        \frac{v}{\sqrt{2}f_{\rm HC}} \, y_u^{\rm SM} & y_U & 0 & 0 \\
        0 & \frac{v}{\sqrt{2}f_{\rm HC}} \, y'_u & y_Q & 0 \\
        0 & 0 & \frac{v}{\sqrt{2}f_{\rm HC}} \, y_d^{\rm SM} & y_D
    \end{array} \right)  \,.
    \label{eq:Y}
\end{equation}
The first matrix, $Y^{\rm SM}$ in Eq.~\eqref{eq:Y} contains the usual Standard Model quark Yukawa couplings, and can be diagonalized through the singular value decomposition with a chiral rotation on the quarks. This procedure introduces a new phase
\begin{equation}
    \bar{\theta}_c 
    = \theta_c - \arg \det Y^{\rm SM} 
    = \theta_{N + 3} - \arg \det \left(y^{\rm SM}_u\,y^{\rm SM}_d\right) \,,
    \label{eq:theta_bar_c}
\end{equation}
where the second relation in \eqref{eq:theta} has been used.
Similarly, the second matrix, $Y^{\rm D}$ can also be diagonalized with a corresponding chiral rotation on the $\SU(N)_D$ quark fields
\begin{equation}
    \bar{\theta}_D 
    = \theta_D - \arg \det Y^D = \theta_{N + 3} + \theta' -  \arg \det Y^D \,,
    \label{eq:theta_bar_D}
\end{equation}
where the first relation in  \eqref{eq:theta} has been used. Since $v \ll f_{\rm HC}$, the phase contribution of $y^{\rm SM}_{u, d}$ and $y'_{u, d}$ to $\arg \det Y^D$ is negligible. In particular, to leading order in $v/f_{\rm HC}$, $Y^D$ is block diagonal, similar to $Y^{\rm SM}$. In fact, it is straightforward to show that
\begin{equation}
    \arg\det Y^D = 2 \arg\det Y_Q + \arg\det Y_U + \arg\det Y_D + {\cal O} \left(\frac{v}{f_{\rm HC}}\right)^4.
\end{equation}
The effective $\theta$ parameters \eqref{eq:theta_bar_c} and \eqref{eq:theta_bar_D} will later be shown to be cancelled by composite axions.

\section{Composite Axion}
\label{sec:compositeaxion}

\subsection{An Invisible Axion from the Hypercolor Sector}
\label{subsec:massless_fermions_hc}

An enlarged color group can be used to obtain a heavy axion. However, it is instructive to first consider the simpler possibility where $\SU(3)_c$ is not modified in the UV, and the only colored hyperfermions are $\chi_c, \bchi_c$, as in the original $\SU(4)/\Sp(4)$ model~\cite{Barnard:2013zea} or the hypercolor models with colored fermions given in~\cite{Ferretti:2014qta}.
If $\psi$ and $\chi_c, \bchi_c$ are massless, then there are two anomalous global symmetries $\U(1)_{\psi}, \U(1)_{\chi}$, with respect to the hypercolor dynamics. The pseudo Nambu-Goldstone bosons\footnote{These Nambu-Goldstone bosons are assumed to arise from hypercolor bilinear condensates. Other four-fermion condensates may also be possible as long as there remains a hypercolor anomaly-free symmetry.} associated with these symmetries, $\sigma_{\psi}$ and $\sigma_{\chi}$, then obtain the hypercolor and QCD anomalous couplings 
\begin{align}
    \Lag \supset 
    \frac{1}{32 \pi^2} \left( a_\psi \, \frac{\sigma_{\psi}}{f_{\rm HC}} + a_\chi \,\frac{\sigma_{\chi}}{f_{\rm HC}}  - \theta_{\rm HC} \right) F_{\rm HC \mu \nu}^{a} \widetilde{F}_{\rm HC}^{a \mu \nu} + 
    \frac{1}{32 \pi^2} \left( b_\chi \frac{\sigma_{\chi}}{f_{\rm HC}} - \theta_c \right) F_{3\mu\nu}^c \widetilde{F}^{c\mu\nu}_3 \,, 
    \label{eq:hyper_axi}
\end{align}
where $a_{\psi, \chi}$ and $b_{\chi}$ are group theoretical factors that depend on the representations of $\psi, \chi_c, \bchi_c$, and $\theta_{\rm HC},\theta_c$ are the hypercolor and QCD $\theta$ parameters, respectively. For simplicity, we also assume a common decay constant $f_{\rm HC} \approx \Lambda_{\rm HC}/4 \pi$ for both $\sigma_{\psi, \chi}$.
From Eq.~\eqref{eq:hyper_axi}, the hypercolor and QCD instantons generate mass terms
\begin{equation}
    \Lag \supset 
    \frac{1}{2} \Lambda_{\rm HC}^4 \left( a_\psi \frac{\sigma_{\psi}}{f_{\rm HC}} + a_\chi \,\frac{\sigma_{\chi}}{f_{\rm HC}}  - \theta_{\rm HC} \right)^2  +  
    \frac{1}{2} \Lambda_c^4 \left( b_\chi \frac{\sigma_{\chi}}{f_{\rm HC}} - \theta_c \right)^2  \,.
    \label{eq:hyper_axi_veff}
\end{equation}
Expanding \eqref{eq:hyper_axi_veff} around the CP-conserving minimum
(i.e. $\sigma_{\psi, \chi} \rightarrow \langle \sigma_{\psi, \chi} \rangle + \sigma_{\psi, \chi}$), leads to the Lagrangian for the physical fluctuations 
\begin{equation}
\label{eq:axionLag_sigma}
    \Lag_{\rm eff} \supset 
    \frac{1}{2} \Lambda_{\rm HC}^4 \left( a_\psi \frac{\sigma_{\psi}}{f_{\rm HC}} + a_\chi \,\frac{\sigma_{\chi}}{f_{\rm HC}}  \right)^2  +  
    \frac{1}{2} \Lambda_c^4 \left( b_\chi \frac{\sigma_{\chi}}{f_{\rm HC}} +  2 \frac{\eta'_{\rm QCD}}{f_{\pi}} \right)^2  \,,
\end{equation}
where the $\eta'_{\rm QCD}$ of QCD has been included with $f_{\pi}$ the QCD pion decay constant, and the quark mass contributions have been ignored\footnote{Note that the electroweak coset may give rise to additional singlets, as in Eq.~\eqref{eq:su5so5ngb}, however these singlets do not receive mass contributions from the hypercolor instanton (similar to the $\eta_{\rm QCD}$ of QCD not receiving QCD instanton contributions).}. Besides $\eta'_{\rm QCD}$, the physical eigenstates resulting from Eq.~\eqref{eq:axionLag_sigma} are
\begin{equation}
    \sigma_{\rm HC} \approx \frac{a_\psi \sigma_{\psi} + a_\chi \sigma_{\chi}}{  \sqrt{a_\psi^2 + a_\chi^2} }
     \,, \qquad 
    \sigma_c \approx \frac{ - a_\chi \sigma_{\psi} +  a_\psi \sigma_{\chi}}{ \sqrt{a_\psi^2 + a_\chi^2}}
    \,,
\end{equation}
where we neglect next-to-leading corrections of order $(\Lambda_c/\Lambda_{\rm HC})^4$, and the small mixing with $\eta'_{\rm QCD}$. Thus $\sigma_c$ corresponds to a hypercolor anomaly free combination of $\U(1)_\psi$ and $\U(1)_\chi$, while $\sigma_{\rm HC}$ corresponds to the orthogonal linear combination. Since we assume no explicit mass terms, the physical masses are determined purely by instanton contributions given by
\begin{equation}
    m_{\sigma_{\rm HC}} \approx \sqrt{a_\psi^2 + a_\chi^2}\, \frac{\Lambda_{ \rm HC}^2 }{ f_{\rm HC} }
    \,, \qquad
    m_{\eta'_{\rm QCD}} \approx 2 \frac{\Lambda_c^2}{f_{\pi}} \,, \qquad 
    m_{\sigma_{c}} = 0
    \,.
    \label{eq:hyper_axi_masses}
\end{equation}
Thus besides $\eta'_{\rm QCD}$, we obtain a heavy mass eigenstate from the hypercolor instanton contributions, while the other eigenstate is massless. If the quark contributions are included, one recovers the standard axion mass relation~\cite{Kim:1984pt,Choi:1985cb}.
The phenomenology of $\sigma_{\rm HC}$ and $\sigma_c$ was previously studied in Refs.~\cite{Cai:2015bss,Belyaev:2015hgo,Cacciapaglia:2019bqz}, where an explicit mass term was assumed for the light state that kinematically allowed  decays to Standard Model gauge bosons through the WZW interaction terms. However in our case, since we forbid explicit mass terms, the $\U(1)_\chi$ symmetry is only explicitly broken by QCD instantons, and $\sigma_c$ can therefore be a composite invisible axion that solves the strong CP problem.

This can also be understood from a UV - IR matching argument. Above the hypercolor confinement scale, the massless hyperfermions $\psi, \chi_c$, and $\bchi_c$ render the hypercolor and QCD $\theta$ terms irrelevant.
Since CP must still be preserved after confinement, the pseudoscalar singlets $\sigma_{\rm HC}, \sigma_c$ must be axions that dynamically cancel the $\theta$ terms below $\Lambda_{\rm HC}$. Indeed, the minimum of the potential in Eq.~\eqref{eq:hyper_axi_veff} precisely occurs at vacuum expectation values of $\sigma_{\rm HC}$ and $\sigma_c$ which eliminates the CP violating terms in \eqref{eq:hyper_axi}.
Thus, hypercolor can be identified with the axicolor dynamics, as in Ref.~\cite{Choi:1985cb}, except that hypercolor also produces a composite Higgs. Note also that the cancellation in Eq.~\eqref{eq:hyper_axi_veff} only occurs if $\psi$ does not have an explicit mass. If the electroweak coset contains a singlet (such as in Eq.~\eqref{eq:su5so5ngb}), a mass term may be desirable to avoid phenomenological constraints, and the misalignment in \eqref{eq:hyper_axi} may be tolerable for sufficiently small $\psi$ masses. Alternatively, the issue can be completely avoided if the electroweak coset does not contain a singlet, and both possibilities deserves further study.
 
The fact that $\sigma_c$ can be identified with the QCD axion requires the decay constant, which depends on the hypercolor confinement scale, to be $f_{\rm HC} \sim 10^9 - 10^{12}$ GeV. From the perspective of the Higgs sector, this makes the theory unnatural. Nevertheless it is interesting that the composite Higgs potential at $f_{\rm HC}$ has a naturally suppressed quartic coupling, that can provide the appropriate boundary condition to radiatively generate the correct Higgs quartic coupling at low energies~\cite{Redi:2012ad}. Although interesting, we will postpone a study of specific composite Higgs models. In any case, the composite axion $\sigma_c$ behaves as the usual invisible axion at low energies, and solves the strong CP problem.

Instead to construct a heavy composite axion, we will make use of the enlarged color sector, with the diagonal color $\SU(N)_D$ originating from this enlarged color. Besides generating the composite Higgs sector, the hypercolor is also responsible for breaking the enlarged color group. In this case, the axion may receive sizeable contributions from $\SU(N)'$ instantons, and thus could be heavy. However since this $\SU(N)'$ color also has a $\theta'$ term, other massless fermions will need to be introduced into the UV theory in order to prevent a misalignment of the strong CP axion solution.

\subsection{A Heavy Axion from Enlarged Color}
\label{subsec:massless_fermions_cut}

For the enlarged color sector to be totally free of the strong CP problem in the UV, the theory must admit massless fermions charged under both $\SU(N + 3)$ and $\SU(N)'$. 
It is possible to construct a solution with heavy axions that are identified with the pseudoscalar bound states of massless hyperfermions, $\chi$, $\bchi$ and $\chi'$, $\bchi'$. In such a solution, one would have to include the hypercolor $\theta$ angle, such as in Eq.~\eqref{eq:hyper_axi_veff}. Instead, for simplicity, we will consider models with extra massless quarks that are hypercolor singlets. A theory with massless colored hyperfermions will be considered in future work.

We begin by introducing a new massless Weyl fermion, $\xi$ charged under $\SU(N + 3)$. For $\SU(N+3)$ to be anomaly free, the representation of $\xi$ must be either real or pseudoreal\footnote{The Witten anomaly~\cite{Witten:1982fp,Wang:2018qoy} for an odd number of Weyl fermions in the pseudoreal representation only occurs for $\SU(2)$.}. 
If the representation is pseudoreal then the bilinear $\xi \xi$ is zero and a $\xi$ mass term is forbidden. 
The pseudoreal representation is the totally antisymmetric $n$-index tensor
of $\SU(2n)$ where $n = 3, 5, \ldots$ is an odd integer. Thus, $\SU(N+3)$ admits pseudoreal representations for $N=2n-3 = 3, 7, 11, \ldots$ 
The smallest pseudoreal representation of $\xi$ is the $\bf 20$ of $\SU(6)$ associated with $N=3$, which was also considered in Ref.~\cite{Gaillard:2018xgk}. In general, under $\SU(3)_c \times \SU(N)_D$ the pseudoreal representation ${\bf A}_{n}$ decomposes as
\begin{equation}
    \xi({\bf A}_{n}) \to 
    \psi_D                      \,({\bf 1} , \, {\bf n_1}) +
    \bpsi_D          \,({\bf 1} , \, {\bf \bar{n}_1}) +
    \psi_c'                     \,( {\bf 3},  {\bf \bar{n}_2} ) +  
	\bpsi_c'         \,  ( {\bf \bar{3}}, {\bf n_2} ) \,,
	\label{eq:pseudo_branching}
\end{equation}
where  $\text{dim}\,\xi = \icol{2n\\n}$, $n_1 = \icol{2n-3\\ n}$ and $n_2 = \icol{2n-3\\ n-1}$.
In particular, $\xi$ gives rise to massless $\SU(N)_D$ fermions as well as massless bifundamental fermions charged under $\SU(3)_c$ and $\SU(N)_D$. The diagonal color group $\SU(N)_D$ actually plays the role of axicolor~\cite{Kim:1984pt, Choi:1985cb}. When $\SU(N)_D$ confines, these fermions can form axicolor bound states, $\psi_D \bpsi_D$ and $\psi_c' \bpsi_c'$, with a linear combination giving rise to a composite axion that will later be shown to be relevant for the strong CP problem.

For $\SU(2n)$ where $n$ is an even integer, the totally antisymmetric  representation, ${\bf A}_{n}$, is real instead of pseudoreal. After the enlarged color group is broken, $\xi$ has a similar breaking pattern to Eq.~\eqref{eq:pseudo_branching}. For $N > 4$ however, there is a smaller real representation, namely the adjoint representation admitted by all values of $N$. Upon the breaking of the enlarged color symmetry, the adjoint state $\xi$ becomes
\begin{equation}
	\xi ({\bf Ad}) \to 
	\lambda_D    \, ( {\bf 1}, {\bf Ad} ) + 
	\lambda_c    \, ( {\bf 8}, {\bf 1}) + 
	\psi_c'     \, ({\bf 3},{\bf \overline{N}}) +  
	\bpsi_c'   \, ({\bf \overline{3}} ,{\bf N}) + 
	\nu'        ( {\bf 1}, {\bf 1} )\,.
	\label{eq:real_branching}
\end{equation}
The difference between $\xi({\bf Ad})$ and $\xi({\bf A}_{n})$ in Eq.~\eqref{eq:pseudo_branching} is the appearance of the adjoint states $\lambda_D$, $\lambda_c$ and a sterile neutrino $\nu'$. Furthermore, the colored $\psi'_c, \bpsi'$ are in the fundamental representation of $\SU(N)_D$, instead of being in a larger representation.   

If $\xi$ is in a real representation, it is possible for $\xi$ to obtain a Majorana mass. This mass is a free parameter of the theory, and will be set to zero. As a result, there is a global $\U(1)_{\xi}$ symmetry. There is the issue of quantum gravitational corrections breaking the $\U(1)_{\xi}$ global symmetry. However, we will assume that the global symmetry can be gauged, embedded into a larger gauge group structure, or the mass term forbidden with a discrete symmetry, such as $\mathbb{Z}_N$ $(N \geq 3)$, and we will not present the details of this additional structure.

Given the two possibilities \eqref{eq:pseudo_branching} and \eqref{eq:real_branching} for a massless $\xi$, we will choose the adjoint representation to present details of the axion mechanism, since the choice of $N$ is less constrained, and also because the resulting structure is generally simpler. For larger values of $n$, the dimension of $\xi(\rep{Ad})$ is $4n^2 - 1$, while the dimension of $\xi(\rep{A}_n)$ grows exponentially with $n$. Nonetheless, the axion mechanism for both representations is the same, and the results for $N=3,7,\dots$ can be interpreted for a pseudoreal representation. When the diagonal color group $\SU(N)_D$ confines at some scale $\Lambda_D,$ the two states $\psi_c' \, (\rep{3}, \brep{N})$ and $\bpsi_c' \, (\brep{3} ,\rep{N}) $ form a bound state. At this energy scale ($\gtrsim$ 10 TeV), the QCD coupling is weak and QCD color can be treated as flavor with respect to $\SU(N)_D$. Just like the QCD chiral symmetry, the global symmetry for diagonal color is then $\U(3)_L \times \U(3)_R$, giving rise to a singlet meson $\eta'_c$, similar to the $\eta'_{\rm QCD}$ of QCD.
Since the $\eta'_c$ constituents $\psi_c', \bpsi_c'$ are massless, $\eta'_c$ can be relevant as a dynamical axion. 
Furthermore, the condensate $\braket{\psi_c' \bpsi'_c}$ spontaneously breaks the $\U(1)_{\xi}$ symmetry. While the $\SU(N)_D$ fermions in \eqref{eq:real_branching} confine at the $\Lambda_D$ scale, the $\lambda_c,\nu'$ fermions can obtain a mass via a four-fermion interaction between $\psi'_c \bpsi'_c$ and $\lambda_c$ or $\nu'$. 
For instance, the adjoint $\lambda_c$ can obtain mass of order $\Lambda_D^3/M_c^2$, where $M_c$ is related to the mass of some heavy, colored bosons, such as the massive gauge bosons in Eq.~\eqref{eq:massgivegbs}. 
Similarly, the sterile neutrino $\nu'$ can obtain a mass of order $\Lambda_D^3/\Lambda_{\rm UV}^2$, where $\Lambda_{\rm UV}\gg \Lambda_D$. These states could lead to experimental signatures (see Section~\ref{subsec:pheno_axions}).

As mentioned in Section~\ref{subsec:massless_fermions_hc}, besides the QCD chiral anomaly, $\eta'_c$ also couples to the chiral anomaly of $\SU(N)_D$, which may inherit another $\theta$ term from $\SU(N)'$. Since the confinement scale $\Lambda_D$ of $\SU(N)_D$ is at least a few orders of magnitude higher than $\Lambda_c$ of QCD, the $\eta'_c$ axion will align with $\SU(N)_D$ and fail to solve the QCD strong CP problem. To avoid this problem, following~\cite{Gaillard:2018xgk}, we include extra massless Weyl fermions $\psi'$ and $\bpsi'$ (i.e.\,with zero bare mass), which transform in the fundamental and antifundamental representation of $\SU(N)'$, respectively. When the enlarged color group is broken, these fermions transform as a corresponding fundamental (antifundamental) under $\SU(N)_D$:
\begin{align}
    &\psi'^{\dagger} \bar{\sigma}^{\mu}( i \partial_{\mu} - G'^a_{\mu} T'^a ) \psi' + 
    \bpsi'^{\dagger} \bar{\sigma}^{\mu}( i \partial_{\mu} + G'^a_{\mu} T'^a ) \bpsi' \notag \\
    &\to \,i \psi'^{\dagger} \bar{\sigma}_{\mu} \partial^\mu \psi' + i\bpsi'^{\dagger} \bar{\sigma}_{\mu} \partial^\mu \bpsi' - 
        G_D^{a \mu} \left( \psi'^{\dagger}  \bar{\sigma}_{\mu} T_D^a \psi' - \bpsi'^{\dagger} \bar{\sigma}_{\mu} T_D^a \bpsi' \right)  + \ldots \,,
\end{align}
where $G_D$ has again been rescaled as in \eqref{eq:gauge_split} and the terms coupling to the massive enlarged color gauge bosons have been omitted. 

The extra massless $\psi',\bpsi'$ fermions extend the $\SU(N)_D$ global symmetry to $\U(4)_L \times \U(4)_R = \SU(4)_L \times \SU(4)_R \times \U(1)_V \times \U(1)_A$. When $\psi' \bpsi'$ and $\psi'_c \bpsi_c'$ form condensates near the $\SU(N)_D$ confinement scale, they spontaneously break the chiral symmetry to $\SU(4)_V \times \U(1)_V$, and give rise to sixteen Nambu-Goldstone bosons:
\begin{equation}
    {\bf 16} = {\bf 8}_c + {\bf 3}_c + {\bf \bar{3}}_c + {\bf 1}_c + {\bf 1}_c \,,
    \label{eq:SU(4)NGB}
\end{equation}
labelled by their respective QCD representations. The colored NGBs obtain radiatively-induced masses from QCD gluon loops of order the symmetry breaking scale~\cite{Choi:1985cb}. The two singlets in~\eqref{eq:SU(4)NGB} correspond to the $\U(1)_A$ generator $T_{16} = \text{diag}(1, 1, 1, 1)/(2\sqrt{2})$ with current $j_{16}$, and one of the broken $\SU(4)_A$ generators, $T_{15} = \text{diag}(1, 1, 1, -3)/(2\sqrt{6})$ with current $j_{15}$. The $\U(1)_A$ symmetry is broken by the $\SU(N)_D$ instantons and is thus anomalous. The divergence of the $j_{16}$ current would be proportional to $F_D^a \widetilde{F}_D^a$, but since $\SU(3)_c$ weakly gauges part of the global symmetry, the $\U(1)_A$ chiral transformation also induces an anomalous QCD contribution. The corresponding singlet state associated with $T_{16}$ is similar to the $\eta'_{\rm QCD}$ of QCD, and since $\Lambda_D \gg \Lambda_{\rm QCD}$ will obtain a mass of order $\Lambda_D$. The second current, $j_{15}$ is not broken by the $\SU(N)_D$ instantons, but is also broken by the QCD instantons. With respect to $\SU(N)_D$, the corresponding singlet state is similar to the $\eta_{\rm QCD}$ in QCD, and would be massless. However, the QCD instantons will generate a mass and thus allow this composite pseudoscalar state to be identified with the invisible axion~\cite{Kim:1984pt, Choi:1985cb}. 

It is convenient to work in a different basis of axial vector currents that separates $\psi'_c$ and $\psi'$ by defining
\begin{align}
    j'^{\mu}_{cA} &= \sqrt{\frac{3}{2}} \left( \sqrt{3} j_{16}^{\mu} + j_{15}^{\mu} \right) 
    = \bpsi'^{\dagger}_c \bar{\sigma}^{\mu} \bpsi_c' + \psi'^\dagger_c \bar{\sigma}^{\mu} \psi'_c
    \equiv f_D\, \partial^\mu\eta_c'\,, \label{eq:jprime_c} \\
    j'^{\mu}_{A}  &= \frac{1}{\sqrt{2}} \left( j_{16}^{\mu} - \sqrt{3} j_{15}^{\mu} \right) 
    = \bpsi'^{\dagger} \bar{\sigma}^{\mu} \bpsi' + \psi'^\dagger \bar{\sigma}^{\mu} \psi' 
    \equiv f_D\, \partial^\mu\eta_D'    \,,
    \label{eq:jprime}
\end{align}
where $f_D$ is the axion decay constant.
If $\psi'_c$ and $\psi'$ were the only massless $\SU(N)_D$ fermions above the axicolor confinement scale that render all $\theta$ angles in the theory unobservable, then below the confinement scale appropriate linear combinations of $\eta'_c$ and $\eta'_D$ can be identified with axions which keep the $\theta$ angles unobservable.

However in the enlarged color model, unlike~\cite{Kim:1984pt, Choi:1985cb}, the massless bifundamental fermions, $\psi'_c, \bpsi'_c$ originate from the $\xi$ fermion, which also contains other massless fermions. The anomalous current must therefore include the rotations of these extra fermions, besides the $\U(1)_A$ chiral rotation of $\psi'_c, \bpsi'_c$ and $\psi', \bpsi'$. The correct anomalous current which couples to the axion and generalizes Eq.~\eqref{eq:jprime_c}, can be obtained by matching the anomaly above and below the enlarged color breaking scale, $\Lambda_{\rm HC}$. Above $\Lambda_{\rm HC}$, there is an anomalous $\U(1)_{\xi}$ symmetry, broken by $\SU(N + 3)$ instantons
\begin{align}
    \partial^{\mu} \left( \xi^{\dagger} \bar{\sigma}_{\mu} \xi \right) 
    = - \frac{1}{16 \pi^2} T(\xi) \, F_{N + 3} \widetilde{F}_{N + 3} 
    = - \frac{1}{16 \pi^2} (N + 3) F_{N + 3} \widetilde{F}_{N + 3}  \,,
    \label{eq:anomaly_uv_ad}
\end{align}
where $T(\xi)=N+3$ for the adjoint representation of the Weyl fermion $\xi$.
Below $\Lambda_{\rm HC}$, where the enlarged color is broken, $\xi$ decomposes to $\lambda_D, \lambda_c, \psi'_c$ and $\bpsi'_c$ as in~\eqref{eq:real_branching}. These states inherit the same charges under $\U(1)_{\xi}$, which leads to the anomalous current
\begin{equation}
    j^{\mu}_{\xi} = 
    \lambda_D^{\dagger} \bar{\sigma}^{\mu} \lambda_D + 
    \lambda_c^{\dagger} \bar{\sigma}^{\mu} \lambda_c + 
    \bpsi'^{\dagger}_c \bar{\sigma}^{\mu} \bpsi_c' + \psi'^\dagger_c \bar{\sigma}^{\mu} \psi'_c  
    \equiv f_D\, \partial^\mu\eta_c'     \,.
\end{equation}
It is straightforward to check that the divergence of this current matches with the total divergence above the $\Lambda_{\rm HC}$ scale in Eq.~\eqref{eq:anomaly_uv_ad}, namely
\begin{align}
    \partial_{\mu} j^{\mu}_{\xi}
    &=  - \frac{1}{16 \pi^2} (N + 3) F_{D} \widetilde{F}_{D} -  
        \frac{1}{16 \pi^2} (N + 3) F_{c} \widetilde{F}_{c} \,.
        \label{eq:N+3anomdivergence}
\end{align}
Note that this matching condition must be satisfied regardless of the enlarged color representation of $\xi$. For example, when $\xi$ is in the pseudoreal representation, $j_{\xi}^\mu$ must include the transformations of not only $\psi'_c, \bpsi'_c$, but also $\psi_D, \bpsi_D$, as given in Eq.~\eqref{eq:pseudo_branching}. The ${\bf 20}$ dimensional representation of $\SU(6)$ is special, since $\psi_D, \bpsi_D$ are singlets. Again the representation of $\xi$ does not matter, because regardless of how $\xi$ decomposes under the broken enlarged color group, these new states still contribute to $j^{\mu}_{\xi}$ in order to match the anomaly. Nonetheless, the $\xi$ adjoint fermion requires some caution, since $\xi$ decomposes to adjoint fermions $\lambda_c,\lambda_D$, under $\SU(3)_c\times \SU(N)_D$. These adjoints may in turn form condensates, similar to the gluino condensate in the $N = 1$ super Yang-Mills theory. In this case, these condensates are $\theta$ dependent~\cite{SHIFMAN1988445}, and thus the $\theta$ parameter is also unphysical. In our model, we will show that $\lambda_c$ will receive a dynamical mass above the TeV scale, and thus a $\lambda_c$ condensate will not be formed. 
In such cases it has been shown that there is an anomaly-free divergence with a massless Nambu-Goldstone boson~\cite{Smilga:1982in, Farrar:1982te, Farrar:1996ye}. 

Next, we include the other anomalous current \eqref{eq:jprime} arising from the massless $\SU(N)'$ fermions with corresponding divergence
\begin{equation}
    \partial_{\mu} j'^{\mu}_A = - \frac{1}{16 \pi^2}  F_D^a \widetilde{F}_D^a \,.
    \label{eq:primeanomdivergence}
\end{equation}
From Eq.~\eqref{eq:N+3anomdivergence} and~\eqref{eq:primeanomdivergence}, it is now possible to form an $\SU(N)_D$ anomaly free current
\begin{equation}
    \partial_{\mu} \left( j^{\mu}_{\xi} - (N + 3) j'^{\mu}_{A} \right) = - \frac{1}{16\pi^2} (N + 3) F_c \widetilde{F}_c \,,
    \label{eq:j'c}
\end{equation}
which generalizes the $j^{\mu}_{15}$ current corresponding to the case where $\psi'_c, \bpsi'_c$ are not embedded into the $\xi$ fermion. The current given in \eqref{eq:j'c} is only broken by QCD instantons, and shows that there still is a pseudo Nambu-Goldstone boson in the spectrum, which can be identified with the composite invisible axion. 

The effective Lagrangian terms are of the form $\eta'_c \partial_{\mu} j^{\mu}_{\xi}$ and $\eta'_D \partial_{\mu} j'^{\mu}_{A}$, which gives
\begin{equation}
    \Lag_{\text{eff}} = 
    - \frac{1}{32 \pi^2} \left( 2(N+3) \frac{\eta_{c}'}{f_D} + 2 \frac{ \eta_D' }{ f_D } - \bar{\theta}_D \right) F_D^a \widetilde{F}_D^a  - 
    \frac{1}{32 \pi^2} \left( 2(N+3) \frac{\eta_{c}'}{f_D} - \bar{\theta}_c  \right) F_3^c \widetilde{F}_3^c \,,
    \label{eq:LagaxionFFtilde}
\end{equation}
where we have used the relations \eqref{eq:theta_bar_c}, and \eqref{eq:theta_bar_D} for the effective $\theta$ terms with Yukawa phases. 
To the extent that $N$ is large, the effective potential for these pseudoscalars can be written as~\cite{DiVecchia:1980yfw} 
\begin{equation}
    \Lag_{\text{eff}} = 
          \frac{1}{2} \Lambda_D^4 \left(  2(N+3) \frac{\eta_{c}'}{f_D} + 2 \frac{ \eta_D' }{ f_D }  - \bar{\theta}_D \right)^2 
        +\frac{1}{2} \Lambda_{c}^4 \left( 2(N+3) \frac{\eta_{c}'}{f_D} - \bar{\theta}_c \right)^2 \,.
        \label{eq:effquadLag}
\end{equation}
Since all the constituent fermions are massless, the axion mass squared in \eqref{eq:effquadLag} is determined by the topological susceptibility from the pure Yang-Mills gauge theory~\cite{Witten:1979vv}.
Note that if $\eta'_D=0$, the $\eta'_c$ axion cannot solve the strong CP problem because its minimum would be dominated by the first term in \eqref{eq:effquadLag}. This is why the second axion, $\eta_D'$ (with corresponding fermions $\psi',\bpsi'$) is needed~\cite{Kim:1984pt, Choi:1985cb, Gaillard:2018xgk}. It may appear that including the $\eta'_{\rm QCD}$ contribution at the QCD confinement scale in \eqref{eq:effquadLag}, could play the role of $\eta_c'$. However, the shift of $\eta'_{\rm QCD}$ to the new minimum causes $\bar{\theta}_c$ to appear in the quark masses as a phase and thus the strong CP problem would not be solved\footnote{Of course, this would be a solution to the strong CP problem if, for instance, the up quark were massless.}.

Expanding \eqref{eq:effquadLag} around the CP-conserving minimum
(i.e. $\eta'_{c,D} \rightarrow \langle \eta'_{c,D}\rangle +\eta'_{c,D}$), leads to the Lagrangian for the physical fluctuations 
\begin{equation}
\label{eq:axionLag}
    \Lag_{\rm eff} \supset 
        \frac{1}{2} \Lambda_D^4 \left(  2(N+3) \frac{\eta_{c}'}{f_D} + 2 \frac{ \eta_D' }{ f_D }  \right)^2 + 
        \frac{1}{2} \Lambda_{c}^4 \left( 2(N+3) \frac{\eta_{c}'}{f_D} + 2 \frac{\eta'_{\rm QCD}}{f_{\pi}}  \right)^2 \,.
\end{equation}
After diagonalizing the corresponding mass matrix one obtains a massless axion, together with a heavy, dark axion with mass of order $\Lambda_D$. The massless eigenstate is identified with the usual QCD axion after including corrections from the Standard Model quark masses.
These will arise by adding the contribution from $\eta'_{\rm QCD}$ to $F_3^c \widetilde{F}_3^c$ in Eq.~\eqref{eq:LagaxionFFtilde}, which leads to the mixing term between $\eta'_{\rm QCD}$ and $\eta'_c$ in Eq.~\eqref{eq:axionLag}. A nonzero axion mass is then obtained from the quark mass contribution to the $\eta'_{\rm QCD}$ mass squared.
While this correction to the $\eta'_{\rm QCD}$ mass is negligible, the axion mass is now proportional to the quark mass, and as expected vanishes in the chiral limit. This reproduces the composite axion model of Ref.~\cite{Kim:1984pt, Choi:1985cb} with axicolor identified with the $\SU(N)_D$ gauge group. However, since the QCD gauge group and $\SU(N)_D$ are combined into a larger group structure, there are additional contributions which can increase the QCD axion mass and these will be considered next.

\subsection{Effect of Small Instantons}
\label{subsec:smallinstantons}

It has long been known that if QCD is modified at high energies to become strong then small (UV) instantons can give rise to sizeable contributions to the axion mass~\cite{Holdom:1982ex,Dine:1986bg,Flynn:1987rs}. In our particular setup, where QCD is embedded into the enlarged color group $\SU(N+3)\times \SU(N)'$, it is possible for the $\SU(N)'$ instantons at a sufficiently large breaking scale ($\Lambda_{\rm HC}$) to induce a large axion mass~\cite{Gaillard:2018xgk}. There are also $\SU(N+3)$ instantons, but the small $\SU(N)'$ instantons can dominate provided that $\alpha'(\Lambda_{\rm HC})\gg \alpha_{N+3}(\Lambda_{\rm HC})$.

Under this assumption the contribution from the small $\SU(N)'$ instantons induces a scale $\Lambda_I$ determined by an integral over the instanton size $\rho$. Since the $\SU(N)'$ gauge symmetry is spontaneously broken at the scale $\Lambda_{\rm HC}$ only instantons of size $\rho\leq 1/\Lambda_{\rm HC}$ are important~\cite{tHooft:1976snw}. The leading order contribution, which follows the minimal recipe~\cite{Shifman:1979uw}, ignoring higher order terms suppressed by group theoretical factors, is given by\footnote{Hypercolor is treated as a flavor of $\SU(N)'$, or to be more exact, a subgroup of the global flavor group $\SU(n_{\chi} + 1)_{\chi',\psi'} \times \SU(n_{\chi} + 1)_{\bchi'^\dagger,\bpsi'^\dagger}$. Since the instanton only breaks the anomalous $\U(1)_A$ symmetry, the effective fermion interaction must be invariant under the global flavor group, and thus is hypercolor invariant. In \eqref{eq:LamI}, this is clear as the fermion condensate product arises from the $(n_{\chi} + 1) \times (n_{\chi} + 1)$ determinant~\cite{tHooft:1976snw}.}
\begin{equation}
\label{eq:LamI}
    \Lambda_I^4 \approx  \int_{1/M_{\rm P}}^{1/\Lambda_{\rm HC}} \frac{\dd \rho}{\rho^5} D[ \, \alpha'(1/\rho) \,  ] 
                         ~ \left( - \frac{4 \pi^2}{N}  \rho^3 \braket{\psi' \bpsi'} \right) 
                           \prod_{i = 1}^{n_{\chi}}
                           \left( - \frac{4 \pi^2}{N}  \rho^3 \braket{\chi'_{i} \bchi'_{i}} \right)
                           \frac{1}{(4\pi)^6}
                     \prod_{k=1}^3
                     y_u'^{\,k}y_d'^{\,k}\,,
\end{equation}
where  $\alpha'(1/\rho)$ is the $\SU(N)'$ coupling evaluated at the scale $1/\rho$, $y'^k_u, y'^k_d$ are the Yukawa couplings of the primed sector and the Planck scale, $M_{\rm P}$ represents the UV cutoff of the model, and $n_{\chi}$ is the number of $\chi'_i,\bchi'_i$ fermions.
Note that the $\SU(N)'$ fermions $\psi',\bpsi',\chi'_i,\bchi'_i$ contribute to  \eqref{eq:LamI} via their respective $\SU(N)_D$ and hypercolor condensates, $\braket{\psi' \bpsi'}$ and $\braket{\chi'_i \bchi'_i}$, where for $\chi'_i\bchi'_i$, we have neglected the contribution from $m_{\chi'} \ll \Lambda_{\rm HC}$ \footnote{Note that $\braket{\chi'_i \bchi'_i}$ represents the trace separated out from $R'$ in Eq.~\eqref{eq:chi_Delta_R}.}. 
This new contribution from the hypercolor condensate $\braket{\chi'_i \bchi'_i} \sim -\Lambda_{\rm HC}^3$, is much larger than the $\SU(N)_D$ condensate $\braket{\psi' \bpsi' } \sim -\Lambda_D^3$, which was previously considered in Ref.~\cite{Gaillard:2018xgk}.
If $y_{u,d}'$ become sufficiently small, other diagrams involving \eqref{eq:veccoupling} and \eqref{eq:yukawacoupling} can also contribute to \eqref{eq:LamI}~\cite{Gaillard:2018xgk}.
However, we will assume that \eqref{eq:LamI} is the dominant contribution.

The instanton density $D[\alpha']$ is defined to be
\begin{equation}
    D[ \alpha'(1/\rho) ] =  e^{-\frac{2 \pi}{\alpha'(1/\rho)}} \left( \frac{2 \pi}{\alpha'(1/\rho)} \right)^{2N}  c_I(N, N_f) \,,
    \label{eq:instdensity}
\end{equation}
where
\begin{equation}
c_I(N, N_f) = \frac{2^{2(1-N)}}{\pi^2} \frac{e^{-c(1)+2c(\frac{1}{2})(2 -N +N_f)} }{(N - 1)! \, (N - 2)!}\,,
\end{equation}
with $c(\frac{1}{2})=0.145873$, $c(1)=0.443307$ (see Ref.~\cite{tHooft:1976snw}) and $N_f = 7 + n_{\chi}$ is the number of $\SU(N)'$ Dirac fermions above the $\Lambda_{\rm HC}$ scale for the hyperfermion representation, $R_\chi$. The running coupling $\alpha'(\mu)$ is given by
\begin{equation}
    \frac{2\pi}{\alpha'(\mu)} = \frac{2\pi}{\alpha'(\Lambda_{\rm HC})} + b' \ln \frac{\mu}{\Lambda_{\rm HC}}\,, 
    \label{eq:alpha_primed}
\end{equation}
where the one-loop $\beta$-function coefficient $b' = (11 N - 2 N_f)/3.$ 

While the integral in Eq.~\eqref{eq:LamI} can be easily numerically evaluated, it is possible to obtain a simple analytic expression. Assuming $\alpha'(\Lambda_{\rm HC}) \ll 2\pi$, the integral in Eq.~\eqref{eq:LamI} admits the closed form expression: 
\begin{align}
    \Lambda_I^4 \simeq \Lambda_D^3 \Lambda_{\rm HC}  \, e^{- \frac{2 \pi }{ \alpha'(\Lambda_{\rm HC}) }}
    \left( \frac{2 \pi}{\alpha'(\Lambda_{\rm HC})} \right)^{2N} 
    \left( \frac{4\pi^2}{N} \right)^{n_\chi + 1} 
    \frac{ c_I(N, N_f) }{ (3 n_{\chi} + b' - 1) }
    \prod_{k=1}^3 \frac{y_u'^{\, k}}{4 \pi} \frac{y_d'^{\, k}}{4 \pi} (1 + \delta_{1/M_P}) \,,
    \label{eq:LamIanalytic}
\end{align}
where
\begin{equation}
    \delta_{1/M_P} \propto \frac{N b'}{\pi} \, \frac{\alpha'(\Lambda_{\rm HC})}{(\ln( M_{\rm P}/\Lambda_{\rm HC}))^{2N}} - \left( \frac{\Lambda_{\rm HC}}{M_{\rm P}} \right)^{3 n_{\chi} + b' - 1 } \,,
\end{equation}
is the contribution from instantons of size $1/M_P$. Since $\Lambda_{\rm HC}\ll M_P$, these contributions are negligible and the integral \eqref{eq:LamI} is dominated by instantons of size $1/\Lambda_{\rm HC}$. The approximate expression \eqref{eq:LamIanalytic} then shows that the instanton contribution can be sizeable provided $\alpha'(\Lambda_{\rm HC})$ is not too small, and the scales $\Lambda_D$, $\Lambda_{\rm HC}$ are as large as possible.

The $\eta'_D$ is a bound state consisting of the $\psi' \bpsi'$ fermions, which are charged under $\SU(N)'$ above the $\Lambda_{\rm HC}$ scale. Thus $\eta'_D$ will receive mass contributions proportional to $\Lambda_I^2$ from the $(1/\Lambda_{\rm HC})$ small instantons\footnote{If the hyperfermions $\chi', \bchi'$ are massless, then the $\sigma'$ also receives a mass contribution from $\Lambda_I$. In this case $\eta'_D$ in \eqref{eq:axionLagwSSI} is replaced by a linear combination of $\eta'_D$ and $\sigma'$, and the orthogonal combination is massless. In our model, since $m_{\chi} \Lambda_{\rm HC}$ is heavier than $\Lambda_D^2$ and $\Lambda_I^2$, the dynamical axion masses are instead determined by $\Lambda_D$ and $\Lambda_I$.}. Therefore, the full effective axion potential is
\begin{align}
    \Lag_{\text{eff}} = 
        \frac{1}{2} &\,\Lambda_D^4 \left( 2 \frac{ \eta_D' }{ f_D } + 2 (N + 3) \frac{\eta_{c}'}{f_D} - \bar{\theta}_{D}  \right)^2 + 
        \frac{1}{2} \Lambda_{c}^4 \left( 2(N + 3) \frac{\eta_{c}'}{f_D} - \bar{\theta}_{c} \right)^2 \notag\\
        &+\frac{1}{2}  \,\Lambda_I^4 \left( 2 \frac{ \eta_D' }{ f_D } - \bar{\theta}'  \right)^2 \,,
        \label{eq:axionLagwSSI}
\end{align} 
where the $\bar{\theta}'$ contribution in the last term of \eqref{eq:axionLagwSSI} arises from including $e^{i \bar{\theta}'}$ in the instanton density \eqref{eq:instdensity}.
The shift from $\theta'$ to $\bar{\theta}'$ takes into account the rotation of $\SU(N)'$ fermions that have explicit masses, $\bar{q}', u'$ and $d'$, which must be rotated to eliminate the phases in their respective Yukawa coupling matrices $Y^{\rm SM}$ and $Y^D$. To relate $\bar{\theta}'$ to $\bar{\theta}_D$ and $\bar{\theta}_c$, let us assume that the massless $\psi', \bpsi'$ are not present in the theory. In this case, $\bar{\theta}'$ is the only physical $\theta$ angle in the UV. Below $\Lambda_{\rm HC}$, the physical $\theta$ angle (when there are no $\psi', \bpsi'$ fermions) is $\bar{\theta}_D - \bar{\theta}_c$, due to the existence of the anomalous current $\partial_{\mu} j'^{\mu}_{cA}$, which shifts $\theta_D$ and $\theta_c$ equally, as shown in Eq.~\eqref{eq:N+3anomdivergence} (a special case with only bifundamental fermions is given in~\cite{Karasik:2019bxn}).  
Thus, one must have
\begin{equation}
    \bar{\theta'} = \bar{\theta}_D - \bar{\theta}_c = \theta' - \arg\det Y^D + \arg\det Y^{\rm SM} \,.
    \label{eq:theta_bar_prime}
\end{equation}
Note that Eq.~\eqref{eq:theta_bar_prime} can be checked explicitly at leading order in $v/f_{\rm HC}$. This is done by collecting the phases in $\bar{q}', u'$ and $d'$ used to rotate the $Y^{\rm SM}$ and $Y^D$ phases away in Eq.~\eqref{eq:theta_bar_c},~\eqref{eq:theta_bar_D}.

With massless fermions $\psi', \bpsi'$, the phase $\bar{\theta}'$ becomes unphysical in the UV. Below $\Lambda_{\rm HC}$, this is matched by the shift symmetry of $\eta'_D$ in Eq.~\eqref{eq:axionLagwSSI} that relaxes $\bar{\theta}'$ to zero. This is evident from the axion VEVs that minimize the potential in~\eqref{eq:axionLagwSSI}
\begin{align}
    \braket{\eta'_c} &= \frac{1}{2(N + 3)} \bar{\theta}_c f_D\,, \quad \\
    \braket{\eta'_D} &= \frac{1}{2} \left( \bar{\theta}_D - \bar{\theta}_c \right) f_D = \frac{1}{2} \bar{\theta}' f_D \,,
\end{align}
which exactly cancels the $\theta$ terms in Eq.~\eqref{eq:LagaxionFFtilde}. In other words, introducing the $\SU(N)'$ small instanton contribution does not disturb the original strong CP solution, while simultaneously providing a new mass contribution to the composite axions.  
The physical masses of the fluctuations around these minima are then determined by diagonalizing the mass matrix that is obtained from
\begin{equation}
    \Lag_{\rm eff} \supset 
        \frac{1}{2} \Lambda_D^4 \left(  2(N+3) \frac{\eta_{c}'}{f_D} + 2 \frac{ \eta_D' }{ f_D }  \right)^2 + 
        \frac{1}{2} \Lambda_{c}^4 \left( 2(N+3) \frac{\eta_{c}'}{f_D}+ 2 \frac{\eta'_{\rm QCD}}{f_{\pi}}   \right)^2 +
        \frac{1}{2} \Lambda_I^4 \left( 2 \frac{ \eta_D' }{ f_D }    \right)^2  \,,
    \label{eq:axionmassmatrix}
\end{equation}
which arises by including the small instanton contributions to Eq.~\eqref{eq:axionLag}.
Thus, depending on $\Lambda_D$ and $\Lambda_{\rm HC}$, the axion and the dark axion could be at the TeV scale. To determine the range of axion masses, we next consider the possible values of the strong coupling scales, $\Lambda_D$ and $\Lambda_{\rm HC}$.
\begin{figure}[H]
    	\centering
     	\includegraphics[width = 0.9 \textwidth]{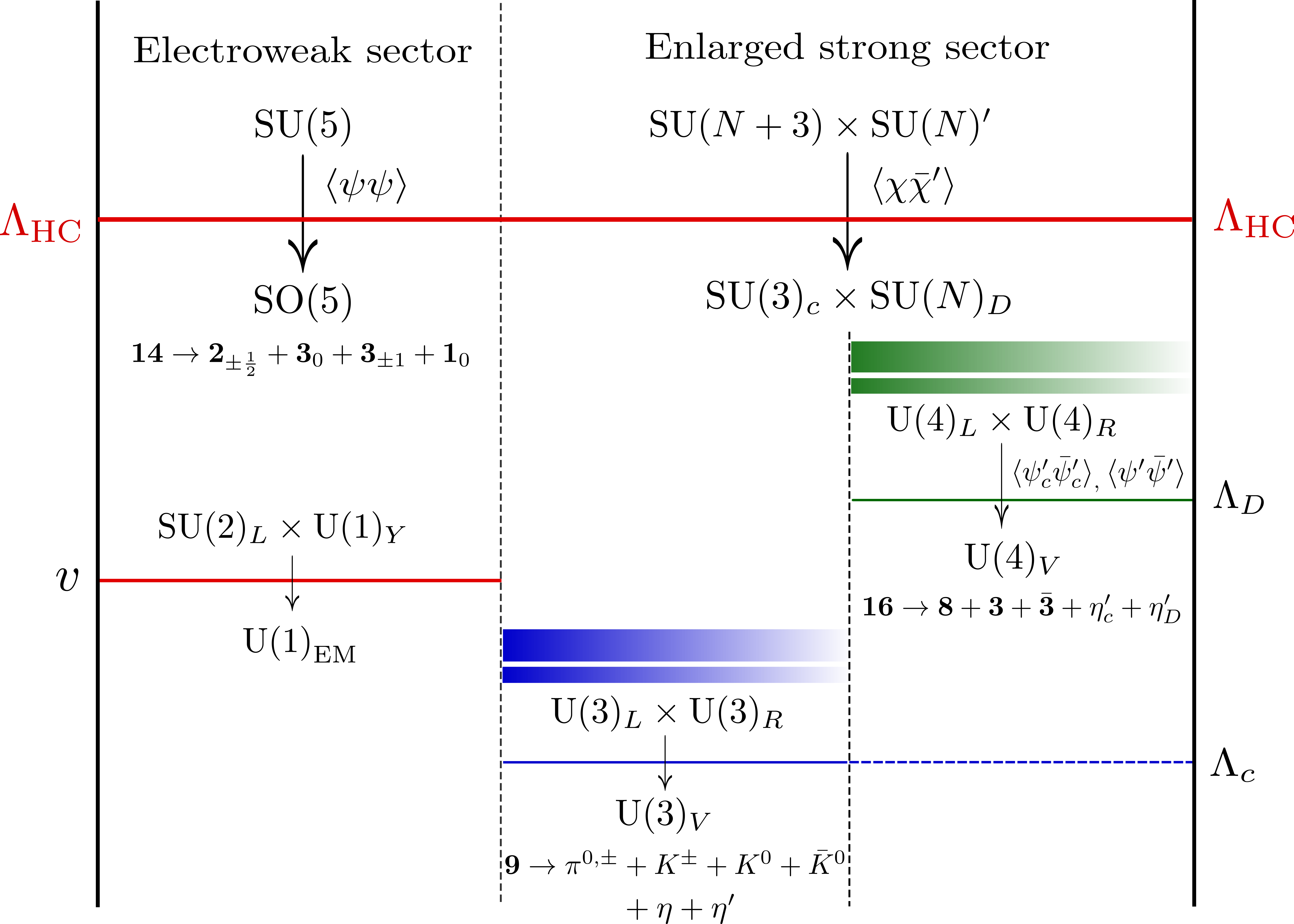}
     	\caption{A schematic diagram showing all the relevant mass scales in the model. Both the electroweak and the enlarged strong sector contain fermions charged under the hypercolor group $\Sp(2N_{\rm HC})$. The green bands around $\Lambda_D$ indicate bound states of the diagonal color group $\SU(N)_D$, while the blue bands around $\Lambda_c$ represent the QCD bound states. The axions $\eta_{c,D}'$ arise from the strong dynamics at  $\Lambda_D$.}
     	\label{fig:scales}
    \end{figure}

\subsection{Axion Mass Scales}
\label{subsec:axionmass}

The axion mass scales are determined from the hypercolor strong scale, $\Lambda_{\rm HC}$. Given the fixed particle content of the model, we can calculate the scale of the condensates $\braket{\psi \psi}$ and $\braket{\chi \bchi'}$, which correspondingly triggers the \SU(5) global symmetry breaking and the enlarged color symmetry breaking, respectively. The massless bound states associated with this global symmetry breaking pattern eventually determine the Higgs and axion mass scales. For example, as has been extensively studied, the Higgs is a pseudo-Nambu-Goldstone boson which eventually develops a VEV and breaks electroweak symmetry. The Standard Model fermion masses then arise from partial compositeness~\cite{Panico:2015jxa}.

Instead for the colored sector, once the condensate $\braket{\chi \bchi'}$ breaks $\SU(N + 3) \times \SU(N)' \to \SU(3)_c \times \SU(N)_D,$ the massless particle content dictates the running of the $\SU(3)_{c,D}$ gauge couplings. This in turn determines the diagonal color confinement scale, $\Lambda_D$ and the corresponding axion bound state mass scale (around $f_D \approx \Lambda_D/(4\pi)$). The instanton contribution, $\Lambda_I$ to the effective potential is then determined from $\Lambda_D$, $\Lambda_{\rm HC}$ and the massless particle content using Eq.~\eqref{eq:LamI}. The physical axion masses can then be obtained by diagonalizing the mass matrix obtained from Eq.~\eqref{eq:axionmassmatrix}. 

For a given scale $\Lambda_D$, how is $\Lambda_{\rm HC}$ constrained? Assuming $\braket{\chi{\bchi'}} \approx -\Lambda_{\rm HC}^3$, the enlarged color group is broken at the hypercolor confinement scale $\Lambda_{\rm HC}$ and the gauge couplings satisfy \eqref{eq:alphaprime}. Below $\Lambda_{\rm HC}$, the QCD coupling runs differently above and below $\Lambda_D$ due to the different fermion content. The one loop expressions are given by
\begin{equation}
	\frac{2\pi}{ \alpha_c (\mu) }   =    
\begin{cases}	
 \frac{2\pi}{ \alpha_c (m_t) } + b_{c1}\ln \frac{\mu}{m_t}\,, \qquad\quad m_t \leq \mu \leq \Lambda_D\,,\\
	\frac{2\pi}{ \alpha_c (\Lambda_{\rm HC}) } + b_{c2}\ln \frac{\mu}{\Lambda_{\rm HC}}\,, \qquad \Lambda_D \leq \mu \leq \Lambda_{\rm HC}\,,
	\end{cases}
	\label{eq:alphaC}
\end{equation}
where $b_{c1},b_{c2}$ are one loop $\beta$-function coefficients, $m_t \simeq 173$ GeV and $\alpha_c (m_t) \simeq 0.1$.
The diagonal color confinement scale is approximately given by:
\begin{equation}
    \Lambda_D\approx \Lambda_{\rm HC} \, e^{-\frac{2\pi}{b_D\alpha_D(\Lambda_{\rm HC})}}\,,
	\label{eq:alphaD}
\end{equation} 
where $b_D$ is the one-loop $\beta$-function coefficient.

Next we determine the $\beta$-function coefficients $b_{c1}, b_{c2}$ and $b_D$. We assume that $\xi$ transforms as an adjoint and use the fermion content listed in Table~\ref{tab:quark_content}. Below $\Lambda_D,$ only the Standard Model quarks contribute to the running
and for six quark flavors this gives the usual Standard Model value $b_{c1} =7$.
However above $\Lambda_D$, there are an additional $N$ Dirac states, $(\psi_c'$,  $\bpsi_c'^\dagger)$, which transform as a fundamental under QCD, and one adjoint Majorana fermion, $\lambda_c$, which gives
$b_{c2} =5 - \frac{2}{3} N$.
Similarly, the $\SU(N)_D$ $\beta$-function coefficient is determined from the four Dirac states, $(\psi_c'$,  $\bpsi_c'^\dagger)$ and $(\psi', \bpsi'^\dagger)$ transforming as a fundamental, and one adjoint Majorana fermion, $\lambda_D$, to give $b_D=3 N - \frac{8}{3}$.
These $\beta$-function coefficients change if $\xi$ transforms in a pseudoreal representation, and can be easily obtained by using the fermion content in Eq.~\eqref{eq:pseudo_branching}.

\begin{figure}[H]
    	\centering
    	\begin{subfigure}[t]{0.6\textwidth}
    	    \includegraphics[width=\textwidth]{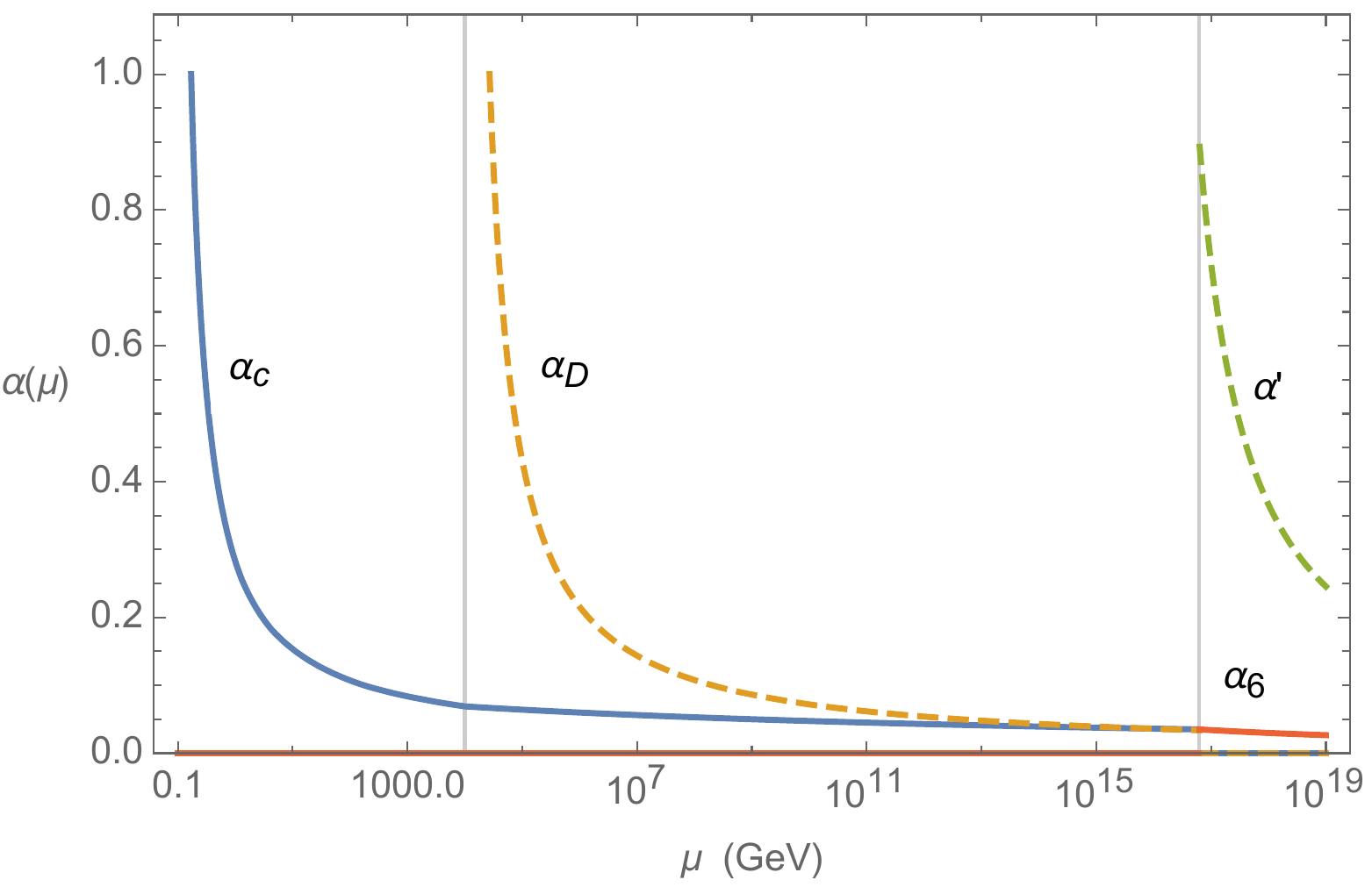}
    	    \subcaption{}
     	\end{subfigure}
     	\begin{subfigure}[t]{0.6\textwidth}
            \includegraphics[width=\textwidth]{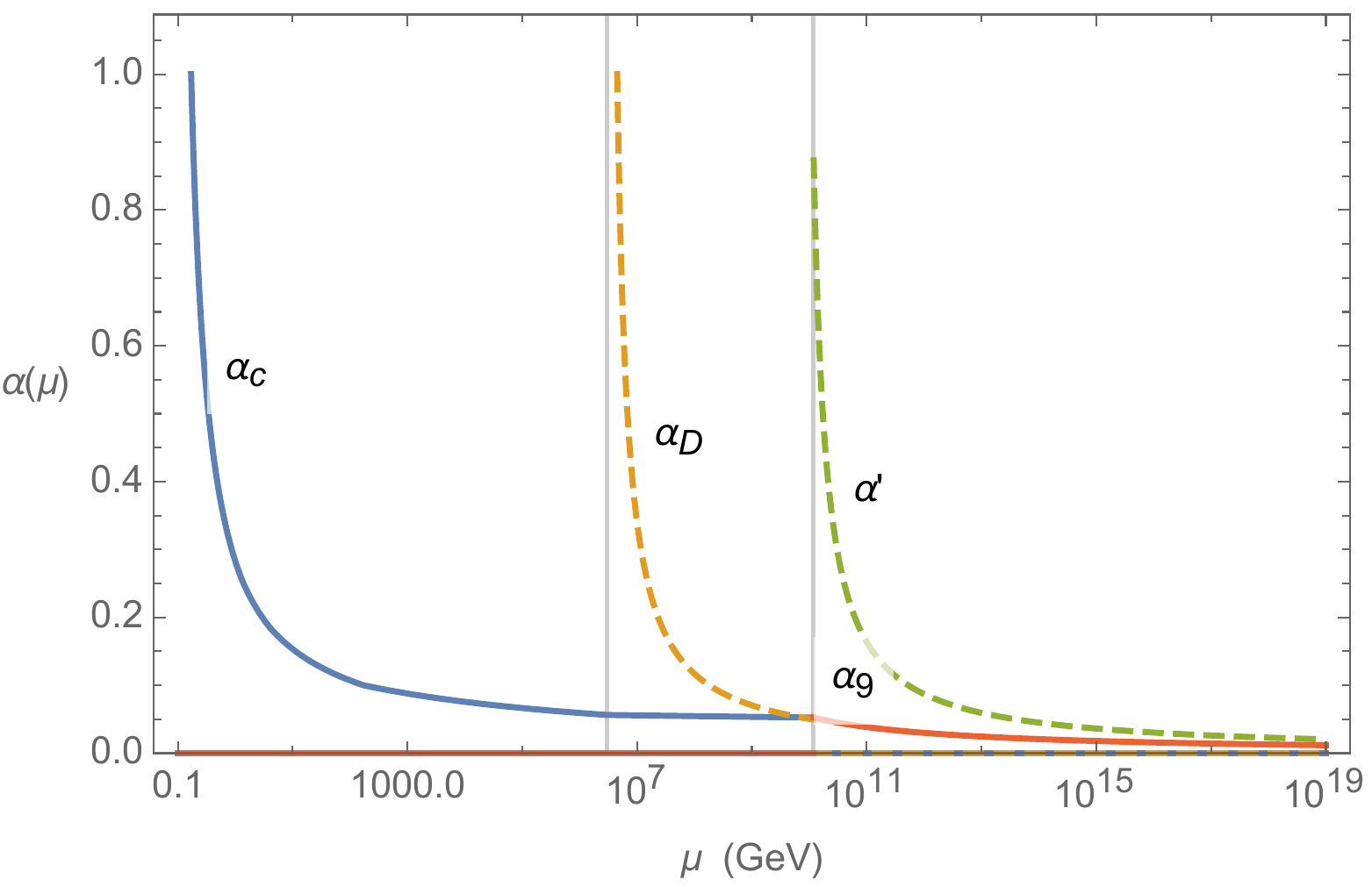}
            \subcaption{}
     	\end{subfigure}
     	\caption{The gauge couplings of the various gauge groups as a function of the renormalization scale. In (a) we assume a hypercolor group $\Sp(4)_{\rm HC}$, $N=3$,  $\Lambda_{\rm HC}\simeq 6.3\times 10^{16}$ GeV and $\alpha'(\Lambda_{\rm HC})=0.9$, giving rise to $\Lambda_D=10$ TeV using \eqref{eq:lamHCratio}, while in (b) we have the hypercolor group $\SU(4)_{\rm HC}$, $N=6$, $\Lambda_{\rm HC}\simeq 1.2\times 10^{10}$ GeV and $\alpha'(\Lambda_{\rm HC})=0.9$, giving rise to $\Lambda_D=3000$ TeV using \eqref{eq:lamHCratio}. }
     	\label{fig:Running}
    \end{figure}

Using these $\beta$-function coefficient values
in (\ref{eq:alphaC}) and (\ref{eq:alphaD}), together with (\ref{eq:alphaprime}), leads to the relation
\begin{equation}
\label{eq:lamHCratio}
    \ln \frac{\Lambda_{\rm HC}}{\Lambda_D} = \frac{1}{11 N - 23} \left[ 6\pi \left( \frac{1}{\alpha_c (m_t)} + \frac{1}{\alpha' (\Lambda_{\rm HC})} \right) + 21 \ln \frac{\Lambda_D}{m_t} \right]\,.
\end{equation}
Thus by specifying the diagonal color scale, $\Lambda_D$, the hypercolor confinement scale, $\Lambda_{\rm HC}$ can be determined from \eqref{eq:lamHCratio}.
Notice that as $N$ increases, $\Lambda_{\rm HC}$ can be lower for a given value of $\Lambda_D$. Interestingly, the one loop expression \eqref{eq:lamHCratio} shows that in the limit $N \to \infty,$ one obtains $\Lambda_{\rm HC} \to \Lambda_D$. This suggests the possibility of a natural composite Higgs together with a heavy composite axion, with the tradeoff being larger group structure. The gauge couplings and strong scales are shown in Fig.~\ref{fig:Running} for two representative choices of the parameters.

In summary, the factors that control the axion masses are $N$, $\alpha'(\Lambda_{\rm HC})$, the choice of hypercolor gauge group, and the Yukawa couplings, $y'$. As shown in \eqref{eq:lamHCratio}, $N$ and $\alpha'(\Lambda_{\rm HC})$ directly control the ratio $\Lambda_{\rm HC}/\Lambda_D$. For example, when $N = 3$, $\Lambda_{\rm HC}$ may be as high as the Planck scale ($M_{P} = 1.22 \times 10^{19}$ GeV), and for $\alpha'(\Lambda_{\rm HC}) = 0.3 \,(0.7)$, we obtain $\Lambda_{D}= 14.2 \,(45.2)$ TeV. The choice of the hypercolor gauge group determines the number $n_{\chi}$ of $\chi'$ hyperfermions, which then affects the $\alpha'$ running. For $\Sp(2N_{\rm HC})$ with $\chi'$ hyperfermions in the fundamental representation, $\,n_{\chi} = 2 N_{\rm HC}$.
A larger fermion representation reduces the one-loop $\beta$-function coefficient, causing $\alpha'$ to run slower. As a result the $\SU(N)'$ instantons stay relevant for a larger range of energy, which in turn increases $\Lambda_I$. Nonetheless, this integral over small size instantons is highly suppressed. It turns out that  for most of the allowed parameter space, $\Lambda_{I} \le \Lambda_{D}$, with the exception of $N = 3$. However, for $N=3$ (with either real or pseudoreal representations), the $\SU(N)_D$ scale is restricted to values $\Lambda_D\lesssim 50$~TeV for $\Lambda_{\rm HC}\lesssim M_P$.

\begin{figure}[H]
    	\centering
    	\begin{subfigure}[t]{0.45\textwidth}
    	    \includegraphics[width=\textwidth]{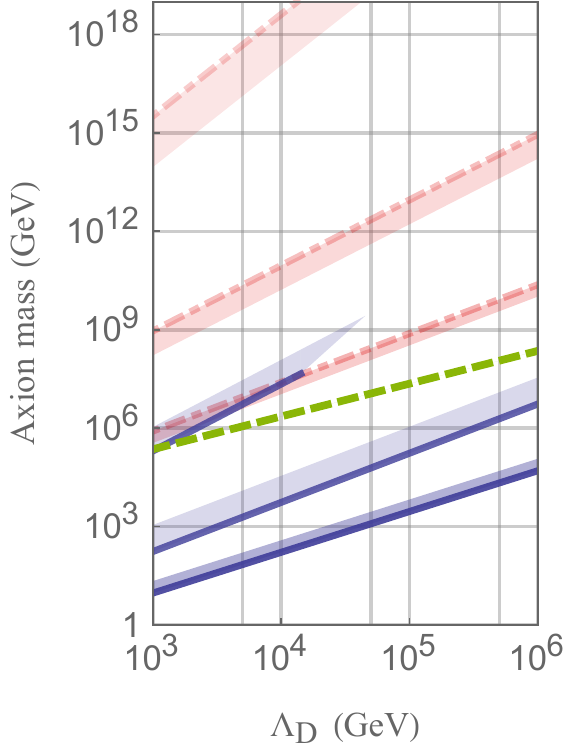}
     	    \subcaption{}
     	    \label{subfig:ma_N}
     	\end{subfigure}
        ~
     	\begin{subfigure}[t]{0.45\textwidth}
            \includegraphics[width=\textwidth]{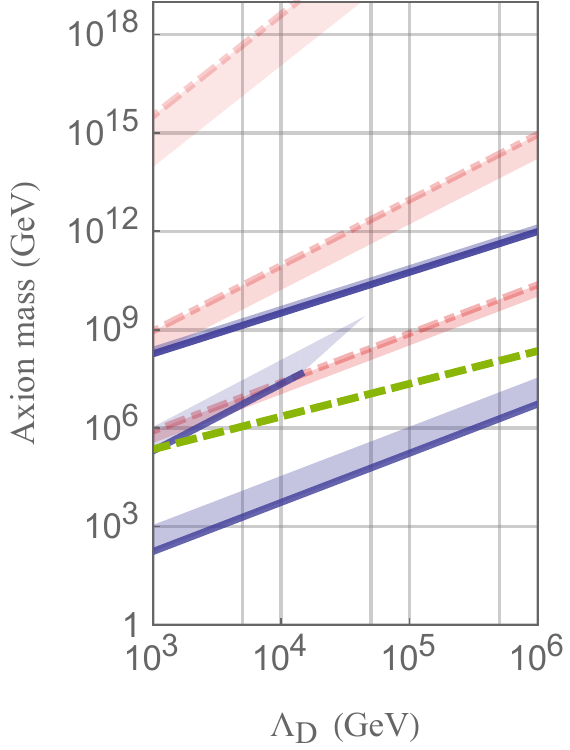}
            \subcaption{}
            \label{subfig:ma_alpha}
     	\end{subfigure}
     	\caption{ The physical axion mass $m_a$ (blue) as a function of $\Lambda_D$, assuming $y_{u,d}'^i=0.1$, $\alpha'(\Lambda_{\rm HC}) = 0.7$, and $N = 3, 4$ and 6 (from top to bottom in the left figure, and middle, bottom, and then top in the right figure). The blue bands indicate the axion mass with $\alpha'(\Lambda_{\rm HC})$ varying from 0.3 to 0.7. The physical dark axion mass $m_{a_D}$ (green, dashed) assuming $\alpha'(\Lambda_{\rm HC}) = 0.7$ for $N = 6$ is also included, which is fairly insensitive to $\alpha'(\Lambda_{\rm HC})$ and $N$. For reference, the red dot-dashed lines represents the $\Lambda_{\rm HC}$ scale determined from Eq.~\eqref{eq:lamHCratio}, for $N = 3, 4, 6$. Like the axion mass bands, the red bands indicate varying $\alpha'(\Lambda_{\rm HC})$. Figure (a) assumes the hypercolor group $\SU(4)_{\rm HC}$, while figure (b) assumes the hypercolor group $\Sp(4)_{\rm HC}$ (for $N = 3, 4$) and $\Sp(20)_{\rm HC}$ (for $N = 6$).  Note that for $N = 6$, the lightest axion actually becomes heavier because $n_{\chi}$ is larger.}
     	\label{fig:ma}
    \end{figure}

The axion masses are plotted in Figure~\ref{fig:ma} for a range of $N$ and $\alpha'(\Lambda_{\rm HC})$ values. The lighter axion mass depends sensitively on $\Lambda_I$. The physical masses are well approximated in the limit $\Lambda_I \lesssim \Lambda_D$ by
\begin{equation}
    m_{a_D}   \approx 2(N +3)\, \frac{\Lambda_D^2}{f_D}\,, \qquad \qquad
    m_{a}\approx 2 \frac{ \Lambda_I^2 }{ f_D }\,, 
    \label{eq:ma_phys}
\end{equation}
where we have ignored the $\eta'_{\rm QCD}$ of QCD, assuming $\Lambda_I\gg \Lambda_{\rm QCD}$. The physical axion mass eigenstates are approximately
\begin{align}
    a   &= \frac{ - (N + 3) \eta'_D + \eta'_c }{\sqrt{ (N+ 3)^2 + 1 }} \approx -\eta'_D + \frac{1}{\sqrt{(N+ 3)}}\eta'_c\,,  \qquad \qquad \notag \\
    a_D &= \frac{ \eta'_D + (N + 3) \eta'_c  }{\sqrt{ (N+ 3)^2 + 1  }} \approx \eta'_c + \frac{1}{\sqrt{(N+ 3)}}\eta'_D\,,
    \label{eq:a_phys}
\end{align}
where $a$ can be identified with the usual QCD axion, while $a_D$ is the $\SU(N)_D$ axion. Note that $a$ aligns mostly with the $\SU(N)_D$ anomaly free current defined in Eq.~\eqref{eq:j'c}.

We consider two limiting cases for the hypercolor confinement scale, where $\Lambda_{\rm HC}$ is as low as $10^{10}$ GeV, or as high as $10^{16}$ GeV.
For the first case, assuming $\Lambda_D = 3000$ TeV, $N = 6$, $\alpha'(\Lambda_{\rm HC}) = 0.9$, $y'^i_{u,d} = 0.025$ and hypercolor group $\SU(4)_{\rm HC}$, one obtains $\Lambda_{\rm HC} \approx 1.2\times 10^{10}$~GeV. This $\Lambda_{\rm HC}$ scale would be relevant for generating a composite Higgs with a vanishing quartic coupling. In this case, the field $\xi$ can only be in the adjoint representation. The axion mass spectrum is
\begin{equation}
    m_{a_D} \approx 6.8 \times 10^8 \text{ GeV } \,, \quad 
    m_a \approx 2.6 \text{ TeV } \,.
    \label{eq:axion_masses_lowhc}
\end{equation}
The dark axion mass is above the $\Lambda_D$ scale, while the axion mass is near the TeV scale, which could potentially be probed at the LHC. Note however that the axion mass depends on the value of the Yukawa couplings, $y'^i_{u,d}$. Smaller values of $y'^i_{u,d}$ will suppress $\Lambda_I$, leading to a lighter axion. This dependence on $y'^i_{u,d}$ also allows for TeV scale axions with other values of $N$ and $\Lambda_{\rm HC}$. 

In the second case, we consider $\Lambda_D = 10$ TeV, $N = 3$, $\alpha'(\Lambda_{\rm HC}) = 0.9$, $y'^i_{u,d} = 0.003$ and hypercolor group $\Sp(4)_{\rm HC}$. For $N = 3$, the field $\xi$ can be in either the real or pseudoreal representation. However for $\xi$ in the real representation, we assume a dynamically generated mass for $\lambda_c$, near the $\Lambda_D$ scale. This will be further discussed in Section~\ref{sec:phenomenology}. In either case, one obtains $\Lambda_{\rm HC} \approx 6.3 \times 10^{16}$~GeV with an axion mass spectrum
\begin{equation}
    m_{a_D} \approx 1.5 \times 10^6 \text{ GeV }\,, \quad 
    m_a \approx 3.1 \text{ TeV }\,.
\end{equation}
In this case, the axion mass is still at the TeV scale, while the dark axion mass is now much below $\Lambda_{\rm HC}$, but still at a higher mass scale.

\subsection{Asymptotic Freedom Condition}
\label{sec:RG_AsymptoticFreedom}

In order for the strong dynamics to confine, the hypercolor gauge coupling should be asymptotically free. This leads to a nontrivial condition on the hyperfermion content $\psi$ and $\Psi_\chi =(\chi, \chi', \bchi, \bchi')$. This can be simply checked by considering the one loop $\beta$-function coefficient of the hypercolor gauge group
    \begin{equation}
        b_{\rm HC} = \frac{11}{3} C_2( {\bf Ad}) - \frac{2}{3} \, n_\psi \, T(R_{\psi}) - \frac{2}{3} \, n_{\Psi_\chi} T(R_{\chi}) \,,
        \label{eq:bHC}
    \end{equation}
where  $C_2( {\bf Ad})$ is the quadratic Casimir invariant, $T(R_{\psi,\chi})$ is the index of the representations $R_{\psi,\chi}$, and $n_{\psi,\Psi_\chi}$ is the number of $\psi,\Psi_\chi$ Weyl hyperfermions.

Consider first the symplectic gauge group $\Sp(2N_{\rm HC})$ where $T( {\bf F}) = \frac{1}{2}$, $T( {\bf Ad}) = C_2( {\bf Ad})=N_{HC} + 1$ and $T({\bf A_2}) = N_{HC} - 1$. In principle there are two choices for the pair of representations $(R_{\psi}, R_{\chi})$.
The first case, $(\bf A_2, \bf F)$, based on the $\SU(5)/\SO(5)$ coset is given in Table~\ref{tab:sp_particle_content}, where $\bf A_2$ could also be replaced with the adjoint representation ${\bf Ad}$. 
The second possibility for $(R_{\psi}, R_{\chi})$ is $(\bf F, \bf A_2)$, which is based on the $\SU(4)/\Sp(4)$ coset for the Higgs sector~\cite{Barnard:2013zea}.
Substituting the hyperfermion content into 
the expression \eqref{eq:bHC} gives
\begin{equation}
    b_{\rm HC} = \begin{cases}
    \frac{1}{3}(19-8N(N_{\rm HC} - 1)-N_{\rm HC}) &\qquad (R_\psi,R_{\chi}) = ({\bf F},{\bf A_2})\,, \\ 
    \frac{1}{3}(15-4N+N_{\rm HC}) &\qquad (R_\psi,R_{\chi}) = ({\bf A_2},{\bf F})\,, \end{cases}
    \label{eq:bHC_Sp}
\end{equation}
where $n_\psi = 5$, $n_{\Psi_\chi}= 4N + 6$, $\text{dim}\,{\bf F}= 2N_{\rm HC}$, and $\text{dim}\,{\bf A_2}=(N_{\rm HC}-1)(2N_{\rm HC}+1)$. 

It is also necessary to check the asymptotic freedom of the $\SU(N)'$ and $\SU(N+3)$ gauge groups, though this will be less stringent than the requirement for $b_{\rm HC}$. For $\SU(N)'$, this requires that the one-loop $\beta$-function coefficient $b'$ to be positive. Using the fermion content in Tables~\ref{tab:sp_particle_content} and \ref{tab:quark_content}, the $b'$ values are
\begin{equation}
    b' = 
     \frac{11}{3} N - \frac{14}{3} - \frac{2}{3} \, n_{\chi'} \,,
    \label{eq:b_primed}
\end{equation}
where $n_{\chi'}={\rm dim} \, R_{\chi'}$ is the number of $\chi', \bchi'$ pairs.
Similarly, the one loop $\beta$-function coefficient, $b_{N+3}$ for $\SU(N+3)$ above the $\Lambda_{\rm HC}$ scale is
\begin{equation}
    b_{N + 3} = 
    3N + 5 - \frac{2}{3} \, n_{\chi} \,,
    \label{eq:b_(n+3)}
\end{equation}
where $n_{\chi}={\rm dim} \, R_{\chi}$ is the number of $\chi, \bchi$ pairs.
Note that \eqref{eq:b_(n+3)} assumes that $\xi$ transforms in the adjoint representation.
When $N=3, 7, \ldots$, $\xi$ may instead transform under the pseudoreal representations listed in Eq.~\eqref{eq:pseudo_branching}.
For $N=3$ one obtains $b_6= 16-\frac{2}{3} \, n_{\chi}$, while for $N=7$,  $b_{10} = \frac{2}{3}(14 - n_{\chi})$.

\begin{figure}[H]
\centering
  \begin{subfigure}[b]{0.46\textwidth}
  \centering
    \includegraphics[width=\textwidth]{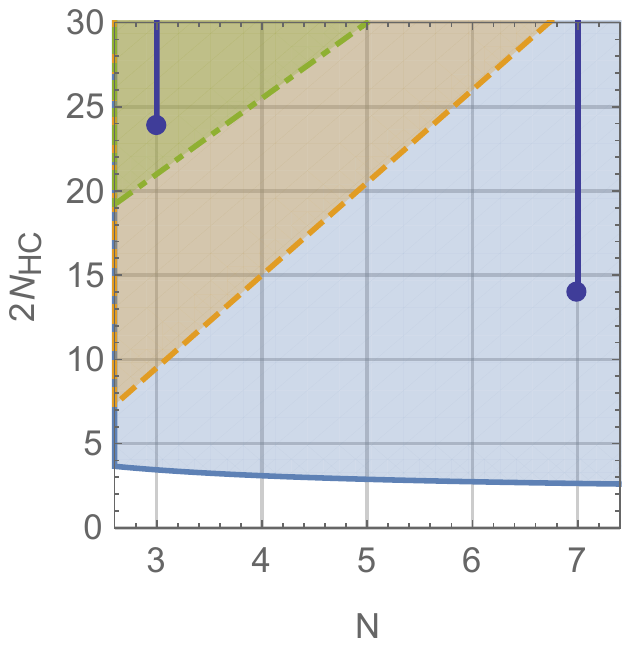}
    \caption{}
    \label{fig:asymptotic_constraint_a}
  \end{subfigure}
  \quad
  \begin{subfigure}[b]{0.46\textwidth}
  \centering
    \includegraphics[width=\textwidth]{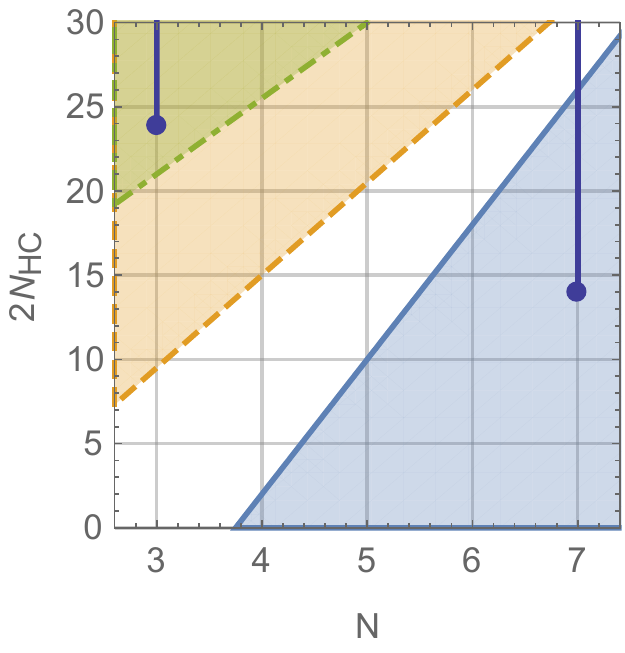}
    \caption{}
    \label{fig:asymptotic_constraint_b}
  \end{subfigure}
  \caption{The allowed region for $N$ and $N_{\rm HC}$ for (a) $(R_{\psi}, R_\chi)= ({\bf F}, {\bf A_2})$, and (b) $(R_{\psi}, R_\chi)=({\bf A_2},{\bf F})$ when the hypercolor group is $\Sp(2N_{\rm HC})$. The shaded blue regions are ruled out by $b_{\rm HC}<0$, while the orange (dashed boundary) and green (dot-dashed boundary) regions are ruled out by $b'<0$ and $b_{N + 3}<0$, respectively. For $\SU(N + 3)$ with $N = 3$ or $7$, the fermion $\xi$ may instead be in the pseudoreal representation, and the allowed limits are indicated by the values of $2 N_{\rm HC}$ less than the dark blue dots.}
  \label{fig:asymptotic_constraint}
\end{figure}

Using the expressions~\eqref{eq:bHC_Sp},~\eqref{eq:b_primed} and \eqref{eq:b_(n+3)}, the results for requiring $b_{\rm HC},b',b_{N+3}>0$ are plotted in Figure~\ref{fig:asymptotic_constraint} for the $\Sp(2 N_{\rm HC})$ hypercolor group.
For the case $(R_\psi,R_{\chi}) = ({\bf F},{\bf A_2})$, the requirement for $\Sp(2N_{\rm HC})$ to be asymptotically free is $N_{\rm HC} < \frac{8 N + 19}{8 N + 1}$.
For $N\geq 3$, there is no solution when $N_{\rm HC}\geq 2$ as can be seen in Figure~\ref{fig:asymptotic_constraint_a}. 
The ${\bf A_2}$ representation for the colored hyperfermions, $\chi$, is just too large for the theory to be asymptotically free.
In the case where $(R_\psi,R_{\chi}) = ({\bf A_2},{\bf F})$, the condition for asymptotic freedom is $N_{\rm HC} > 4 N - 15$. The allowed region for $N$ and $N_{\rm HC}$ is shown in figure~\ref{fig:asymptotic_constraint_b}. In this case we see that for increasing $N_{\rm HC}$ there are values of $N$ for which the theory is asymptotically free, and corresponds to the case considered in Section~\ref{subsubsec:hc_symplectic}. A slight modification of this possibility is to have $(R_{\psi},R_{\chi}) = ({\bf Ad}, {\bf F})$. However in this case asymptotic freedom requires $N_{\rm HC} > 4 N +5$, and there is no value of $N_{\rm HC}$ that allows for an asymptotically free  $\SU(N + 3)$ and $\SU(N)'$ group. 

When the hypercolor gauge group is $\SU(4)_{\rm HC}$ we have $T( {\bf F}) = \frac{1}{2}$, $T( {\bf Ad}) = 4$ and $T({\bf 6}) = 1$. For $(R_{\psi}, R_\chi)=({\bf A_2},{\bf F})$ the condition for $\SU(4)_{\rm HC}, \SU(N+3)$ and $\SU(N)'$ to be asymptotically free is $2<N<7$. This corresponds to the case studied in Section~\ref{subsubsec:hc_unitary}. Note that $N<7$ is also consistent with the QCD coupling remaining asymptotically free above the $\Lambda_D$ scale. Thus we see that if $\chi$ transforms in the fundamental representation $\bf F$, then the hypercolor group remains asymptotically free for a range of $N$. In particular, this favors the $\SU(5)/\SO(5)$ composite Higgs models with either $\Sp(2 N_{\rm HC})$ or $\SU(4)_{\rm HC}$ hypercolor gauge groups. Other hypercolor groups of composite Higgs models classified in Ref.~\cite{Ferretti:2013kya}, may also exist for small $\psi,\chi$ representations.

Finally, we should point out that the asymptotic freedom condition on $\SU(N+3)$ and $\SU(N)'$ is not strictly necessary since these gauge couplings could grow towards the UV. In this case the small instanton contribution in \eqref{eq:LamI} would be UV dominated, and need to be recalculated. Consequently, as seen in figure~\ref{fig:asymptotic_constraint_b}, this would then allow the values $5\leq N_{\rm HC} \leq 12$ in the $N=3$ pseudoreal case, while the axion mass scales for $N_{\rm HC} \leq 4$ arising from small instantons can be obtained from \eqref{eq:LamI} and give similar results to those in Section~\ref{subsec:axionmass}.

\section{Phenomenology}
\label{sec:phenomenology}

\subsection{Axions and \texorpdfstring{$\SU(N)_D$}{SU(N)\_D} Bound States}
\label{subsec:pheno_axions}
The possible lightest states in the model are the dynamical axions with masses determined by $\Lambda_I$, the adjoint fermion $\lambda_c$, and the QCD-like bound states in the dark sector associated with the confinement scale $\Lambda_D$ of the $\SU(N)_D$ strong dynamics. The hypercolor confinement scale is just too large to be phenomenologically relevant for current experiments, although it provides motivation for why the Higgs coupling vanishes at $\Lambda_{\rm HC}\sim 10^{10}$~GeV. Nevertheless, as some of the hypercolor Nambu-Goldstone boson masses are dependent on the hyperfermion masses $m_{\chi, \chi'}$, these states could theoretically be anywhere between $\Lambda_D$ and $\Lambda_{\rm HC}$. Though these states are still likely heavier than the $\SU(N)_D$ lightest bound states, some of the hypercolor Nambu-Goldstone bosons may still affect the ratio $\Lambda_D/\Lambda_{\rm HC}$ because they are charged under QCD and $\SU(N)_D$.

An example of such a light state is the adjoint fermion $\lambda_c$, which only exists in models with $\xi$ in the adjoint representation of $\SU(N+3)$. 
Below $\Lambda_{\rm HC}$, the broken $\SU(N+3)$ gauge bosons can give rise to a four-fermion interaction  $(\lambda_c \lambda_c) (\psi'^{\dagger}_c \bpsi'^{\dagger}_c) + \hc$ between the $\xi$ components. 
At the scale $\Lambda_D$, the $\braket{\psi'_c \bpsi'_c}$ condensate then spontaneously breaks the $\U(1)_{\xi}$ symmetry, and generates\footnote{A similar mechanism in the technicolor context for quarks in the fundamental representation was considered in Ref.~\cite{Dimopoulos:1979es}. The mass may be further enhanced by a factor of $(\Lambda_{\rm HC}/\Lambda_D)^{\gamma}$, where $\gamma$ is the anomalous dimension of the $\psi'_c \bpsi'_c$ operator~\cite{Holdom:1982ex}.} a $\lambda_c$ mass $\sim \Lambda_D^3 / f_{\rm HC}^2$, where the broken gauge boson masses depend on $f_{\rm HC}$ with $f_{\rm HC}\sim \Delta_{n}\lesssim \Lambda_{\rm HC}$.
This $\lambda_c$ mass estimate favors models with a large $\Lambda_D/\Lambda_{\rm HC}$ ratio, i.e. with higher values of $N$.
Thus provided the ratio $\Lambda_D/\Lambda_{\rm HC}$ is not too small, it is possible to dynamically generate a TeV scale mass for $\lambda_c$.
For example, a TeV scale $\lambda_c$ mass can be obtained for the particular 
$\Lambda_D, \Lambda_{\rm HC}$ values considered in Eq.~\eqref{eq:axion_masses_lowhc}, provided
$\Delta_{n} \simeq 0.01 \,\Lambda_{\rm HC}$ (see also Figure~\ref{fig:num_delta}). The $\Lambda_D/\Lambda_{\rm HC}$ ratio can be further increased by including light hypercolor bound states which affect the running of the QCD and $\SU(N)_D$ couplings\footnote{For example, a bifundamental $(\rep{3}, \brep{N})$ complex scalar, changes $b_{c2}\rightarrow b_{c2}-\frac{1}{6}N$ and $b_D\rightarrow b_D-\frac{1}{2}$, and for $N\leq 6$ both QCD and $\SU(N)_D$ couplings remain asymptotically free for $\Lambda_D$ scalar masses. Furthermore, we have checked that the gauge couplings can remain perturbative below $\Lambda_{\rm HC}$ for other colored light scalars that may be experimentally accessible. Nonetheless, a sizeable UV value of the QCD coupling can realize the scenario in Ref.~\cite{Flynn:1987rs}.}.
Alternatively, larger contributions to the $\lambda_c$ mass may arise from a nonperturbatively generated four-fermion coupling or low-lying hypercolor scalar bound states. 
In any case, the adjoint $\lambda_c$ remains a promising signal at the LHC, as it resembles a gluino in supersymmetric models. The $\lambda_c$ decay signal is model dependent, but it is reasonable to expect decays to three jets, via off-shell colored scalars. A recent LHC analysis in this channel gives a gluino mass lower bound of 1.5 TeV~\cite{Sirunyan:2018duw}. 
There is also a singlet fermion component $\nu'$ of $\xi(\rep{Ad})$ which can similarly obtain a mass from higher-dimension terms. It will behave as a sterile neutrino and could be interesting phenomenologically.

If $\Lambda_I > \Lambda_D$, then both axions are several times heavier than the $\SU(N)_D$ scale, which would be the lowest new scale in the model. As a result, the next most promising signals are those arising from the $\SU(N)_D$ bound states, such as dark pions and glueballs. The novel signals in this case consist of the dark pion octet and triplets in Eq.\eqref{eq:SU(4)NGB}, containing the same constituents as the axions. 
With the current LHC limit on exotic pions at 1345 GeV, this translates into the lower bound $\Lambda_D \ge 7 \text{ TeV }$\cite{Aaboud:2019trc} which puts the lighter axion mass, in this case $m_{a_D}$, at $8\pi (N + 3) \Lambda_D \approx 1000$ TeV, far above the current LHC range. 

The more interesting case is when $\Lambda_I < \Lambda_D$, where one of the axion masses, $m_{a}$ can be as low as 1 TeV (see Fig.~\ref{fig:ma}). This heavy axion can potentially be probed in collider experiments. At a collider, the lighter axion can be produced through the digluon channel. The production rate and the signal decay width are dictated by the effective couplings to the QCD and hypercharge anomalies
\begin{equation}
 	\Lag_a =    
  	- \frac{\alpha_c}{4\pi} \frac{N + 3}{\sqrt{(N + 3)^2 + 1}} \frac{a}{f_D} F_{3\mu\nu}^c \widetilde{F}^{c\mu\nu}_3 - 
  	\frac{\alpha_Y}{2\pi}  \frac{ [Y_{\psi'_c}^2 - (N + 3) Y_{\psi'}^2] }{ \sqrt{(N + 3)^2 + 1}} \frac{a}{f_D} B_{\mu\nu} \widetilde{B}^{\mu\nu} \,,
  	\label{eq:axion_coupling}
\end{equation}
where $Y_{\psi', \psi'_c}$ are the hypercharges of $\psi'$ and $\psi'_c$, respectively and $B^{\mu\nu}$ is the $U(1)_Y$ field strength tensor.
The axion production rate is then dependent on the digluon decay rate
\begin{equation}
\Gamma (a \to gg) = \frac{ \alpha_c^2 }{ 8 \pi^3 } \frac{( N + 3)^2}{(N + 3)^2 + 1} \frac{ m_a^3 }{ f_D^2 } \,,
\end{equation}
which is inversely proportional to $f_D$. Thus the signal can be abundantly produced provided $f_D$ is near the TeV scale. Once produced, the axion can decay through the digluon, diphoton, as well as into $Z$ bosons. The corresponding decay rates are~\cite{Gherghetta:2016fhp}
\begin{align}
\Gamma(a \to \gamma \gamma) &= \frac{\alpha_{\rm EM}^2}{2 \pi^3} \frac{ [Y_{\psi'_c}^2 - (N + 3) Y_{\psi'}^2]^2 }{(N + 3)^2 + 1} \frac{m_a^3}{f_D^2} \,,\\
\Gamma(a \to Z \gamma)      &= 2 \tan^2\theta_W\, \Gamma(a \to \gamma\gamma) \,, \\
\Gamma(a \to Z Z)           &= \tan^4\theta_W\, \Gamma(a \to \gamma\gamma)  \,,
\end{align}
where $\alpha_{\rm EM}$ denotes the fine-structure constant and $\theta_W$ is the weak-mixing angle. In this model, just like typical hadronic axions, there is no coupling to leptons and the $W$ bosons. However, new decay channels may arise since there is strong mixing between $\eta'_c$ and $\eta'_D$, allowing for $a$ to decay into $\SU(N)_D$ lighter states, such as dark pions and dark glueballs, assuming these states are kinematically accessible. Specifically, the phenomenological details are dependent on the mass hierarchy of the axion and the lightest states of the dark sector. This is very similar to the hidden valley scenario in twin Higgs models~\cite{Craig:2015pha}, except that instead of the Higgs boson probing the hidden color sector, that role is now played by the heavy axions. 
Nonetheless, if the axion mass is close to $\Lambda_D$, then these decay widths may be suppressed, and thus the main decay channels are still the digluons and diphotons. 

Depending on the free parameters $m_{\chi, \chi'}$, there may still be a light axion in the model such as $\sigma$ and $\sigma'$. For instance, in the limit that $m_{\chi,\chi'}$ vanishes, one can form a combination $\sigma_c$ that couples to an anomaly-free current under $\SU(N)_D$, and thus receives mass contributions only from QCD instantons. This $\sigma_c$ would behave like a standard invisible axion with a decay constant of order $\Lambda_{\rm HC}$, which can be probed in standard invisible axion experiments. The $\sigma_c$ coupling to the gluons is suppressed by $\Lambda_D^2/\Lambda_{\rm HC}^2$, and therefore its most significant coupling is to photons given by
\begin{equation}
	\Lag = \,- n_{\chi} Y_{\chi}^2 \frac{\sigma_c}{f_{\rm HC}} B_{\mu\nu} \widetilde{B}^{\mu\nu} \,,
\end{equation}
where $Y_{\chi}$ is the $\chi$ hypercharge. 
Given that $Y_{\chi}$ is model dependent and not specified, this coupling could be zero leading to a sterile axion-like particle.

\subsection{Cosmology}

The implications and constraints from cosmology are highly dependent on the early history of the universe. It can be shown that there remains a conserved baryon number, and thus the lightest baryon bound state could be stable. If the reheat temperature is above either of the confinement scales then heavy stable particles from the hypercolor or $\SU(N)_D$ sectors may be thermally produced. In the hypercolor case with $\psi$ in the real or pseudoreal representation, it is possible to form a baryon combination $(\psi \psi \psi)$, which could also interact with other baryons, such as the top partner bound state $(\chi \psi \bchi)$. This could cause the bound states to become unstable but a definitive answer would require a detailed study of the hyperbaryon spectrum. 
Nonetheless, it is reasonable to expect that a coupling between these bound states and the Standard Model can be generated when the electroweak symmetry is broken. Thus decays to Standard Model particles would prevent these states from dominating the mass density of the universe. Similarly for the axibaryons, as pointed out in Ref.~\cite{Gaillard:2018xgk}, the enlarged color force between $\SU(N)_D$ and QCD can mediate axibaryon decays. Another issue, shared by the original composite axion models, is domain walls. To avoid these as well as potential stable, heavy baryons, we assume that any domain wall in the model is formed above the inflation scale. Again the details are model dependent, but it would be interesting to study how inflation models could be incorporated into our framework.

Finally, our model has the advantage of admitting several candidates for dark matter. These candidates consist of a light axion-like particle that does not directly couple to QCD, but has a large decay constant determined by the hypercolor scale. Another candidate is the dark photon of the diagonal color that can arise from a possible gauging of the $\U(1)_D$ global symmetry, and although interesting, the phenomenology of this dark photon will not be explored in this paper.

\section{Conclusion}
\label{sec:conclusion}

The Higgs and axion could both arise from the same underlying strong dynamics. In this paper we have given a UV description of a model that produces a composite Higgs and dynamical axions based on the $\Sp(2 N_{\rm HC})$ and $\SU(4)_{\rm HC}$ hypercolor gauge groups. The matter content of the theory consists of colored and uncolored hyperfermions that form condensates at the strong coupling scale $\Lambda_{\rm HC}$. Given the dynamics and hyperfermion matter content, all other scales in the model can be determined from the single scale, $\Lambda_{\rm HC}$.

For the hypercolor group to be asymptotically free, and therefore confine, requires the colored hyperfermions to be in the fundamental representation of the hypercolor gauge group. A particular choice satisfying this condition is a minimal modification of the $\SU(5)/\SO(5)$ composite Higgs model, in which extra colored hyperfermions are added to the theory in order to incorporate an enlarged color sector $\SU(N + 3) \times \SU(N)'$. The hyperfermion condensates then break the enlarged color group to $\SU(3)_c \times \SU(N)_D$. The enlarged color sector also contains massless fermions which are hypercolor singlets and render the $\theta$ angles of the enlarged color group unobservable. When the diagonal color group $\SU(N)_D$ confines at lower energies it produces dynamical axions which preserve the CP symmetry.
Together these massless fermions and axions can eliminate all sources of strong CP violation. A more minimal version of the model is also possible, where QCD is not modified in the UV and the hypercolor dynamics produces an invisible axion, in addition to a heavy axion at the hypercolor confinement scale.

The enlarged color group allows for extra axion mass contributions that arise from small instantons associated with the $\SU(N)'$ gauge group. This leads to TeV scale axions for both the real and pseudoreal representations of the massless enlarged-color fermions, that can be detected at collider experiments via their decays into photons and $Z$ bosons. This possibility exists for both high ($\simeq 10^{15}$ GeV) and low ($\simeq 1000$~TeV) values of the hypercolor confinement scale. Other axion decay channels into $\SU(N)_D$ pions or glueballs may be possible. A dynamical axion arising from the hypercolor dynamics has suppressed couplings to gluons, and instead could couple to photons provided the hyperfermions are charged under $\U(1)_Y$. 
However, besides the TeV scale axions, in the case of a massless fermion in the adjoint representation,  there is also a QCD colored octet fermion in the low energy spectrum. This fermion can receive a dynamically generated TeV scale mass, which can then decay into three jets, providing a smoking gun signal of this scenario.
In addition there are several dark matter candidates including the hypercolor axion, as well as a possible dark photon. 

Our model provides a complete UV description that generates a composite Higgs together with dynamical axions. All mass scales arise from the hypercolor confinement scale which can be as low as 1000 TeV. However, this does imply that the composite Higgs sector must be tuned in order to achieve electroweak symmetry breaking, and other experimental signatures could arise from an ``unnatural" Higgs sector similar to those considered in Ref.~\cite{Barnard:2014tla}.
If the hypercolor strong scale is near the $10^{10-11}$~GeV, then the quartic coupling of the Higgs potential may be generated radiatively as considered in Ref~\cite{Redi:2012ad}.
Furthermore, there may also exist other hypercolor groups with different hyperfermion representations, as well as small instanton contributions arising from UV modifications of QCD that could be incorporated into a composite Higgs framework. Nevertheless, it is clear that an enlarged strong dynamics can extend the range of axion masses, and provides further motivation for experimental searches of axion-like particles.

\section*{Acknowledgments}
We thank Aleksey Cherman, Gabriele Ferretti and Arkady Vainshtein for helpful discussions. This work is supported in part by the Department of Energy under Grant DE-SC0011842 at the University of Minnesota, and T.G. is also supported by the Simons Foundation.

\appendix
\section{Exact one loop effective potential for top partner hyperfermion}
\label{app:A}

In this section, we derive the gap equation for $R, R', \Delta$ and $\tDelta$, with the Yukawa Lagrangian
\begin{align}
        \Lag_{\chi}   &= - \left[ 
                              (R^T)^F_{\x G} ( \chi_F \bchi^G  ) 
                            + (R'^T)^f_{\x g} ( \chi'_f  \bchi'^g ) 
                            + \Delta^{\dagger F}_f (\chi_F \bchi'^f)
                            + \tDelta^{\dagger f}_{\x F} (\bchi^F   \chi'_f) + \hc
                            \right] \notag  \\
                           &- \frac{2 N_{\rm HC}}{ \kappa_{R} } R_F^{\x G} R_G^{\x F} 
                            - \frac{2 N_{\rm HC}}{ \kappa_{R'} } R'^{\x g}_f R'^{\x f}_g 
                            - \frac{2 N_{\rm HC}}{ \kappa_{\Delta} } {\Delta^F}_f  {\Delta^{\dagger f}}_F 
                            - \frac{2 N_{\rm HC}}{ \kappa_{\tDelta} } {\tDelta_F}^{\,\,f} {\tDelta_f}^{\dagger F}\,,
        \label{eq:lag_mix}
\end{align}
by integrating out the hyperfermions. First we wish to derive the exact one-loop effective potential 
\begin{equation}
    V_{\text{eff}} = \,  \sum_{\phi} \frac{2N_{\rm HC}}{\kappa_{\phi}} \tr (\phi \phi^{\dagger}) - 
        \frac{1}{2} \sum_{\phi} \int \frac{d^4k_E}{(2\pi)^4} \, \tr \log \left( 1 + \frac{\phi \phi^{\dagger}}{k_E^2} \right)  + V_{\rm mix}  \,,
    \label{eq:veff_raw}
\end{equation}
where $\phi = R, R', \Delta, \tDelta$. The trace runs over both the enlarged color indices, $F,f$, and the hypercolor indices, which are suppressed. The integrals are performed over Euclidean four momentum, $k_E$ with a UV cutoff $\Lambda_{\rm UV}$, similar to the case for $M^{ab}.$ The first two terms are the standard Coleman-Weinberg potential for $\phi$, and thus we will focus on deriving the last term $V_{\rm mix}$ that mixes different fields. 

First we choose a convention for the enlarged color indices. The fields $\Delta$ and $\tDelta$ can be treated as matrices with the first index associated with the fundamental in $\SU(N+3)$ and the second index with the fundamental in $\SU(N)'$.  Similarly, $R$ is a $(N + 3) \times (N + 3)$ traceless matrix, while $R'$ is a $N \times N$ traceless matrix. For all of these auxiliary fields, the upper indices are fundamental, while the lower indices are anti-fundamental. Thus repeated indices between different fields can be summed only if they have opposite representations. The interacting vertices are
\begin{equation}
    \chi^{\dagger F} R_F^{\x G} \bchi^{\dagger}_G  \,, \qquad
    \chi'^{\dagger f} R'^{\x g}_f \bchi'^{\dagger}_g  \,, \qquad
    \chi^{\dagger F} \Delta_{F}^{\x f} {\bchi'^{\dagger}}_f\,, \qquad
    \bchi^{\dagger}_F \tDelta^F_{\x f} \chi'^{\dagger f} \,. 
    \label{eq:chi_int}
\end{equation}
Note that Eq.~\eqref{eq:lag_mix} and \eqref{eq:chi_int} include some assumptions about the scalar fields. The fields $R, R'$ can be chosen to be real, since if $R$ is complex we can simply redefine $(R + R^{\dagger})_F^{\x G}$ as the new $R_F^{\x G}$. This follows from the fact that $R^{\dagger} (R'^{\dagger})$ has the same index structure as $R (R')$, since both are in the adjoint representation of $\SU(N+3) (\SU(N)')$. 
Similarly, $\Delta_F^{\x f}$ is in the same representation as $\tDelta^{* \, f}_F$, and thus $\tDelta^{* \, f}_F$ share the same coupling to $\chi, \chi'$ with $\Delta_F^{\x f}$. These couplings are not present in Eq.~\eqref{eq:lag_mix}, \eqref{eq:chi_int} since we can redefine $(\Delta + \tDelta^*)_F^{\x f}$ as $\Delta_F^{\x f}$ . Ignoring the momenta and the integral over momentum space, the lowest order Feynman diagrams give
\begin{equation}
    \tr \left[ \Delta R' \Delta^{\dagger} R + \tDelta R'^{T} \tDelta^{\dagger} R^T + \hc \right] \,.
    \label{eq:delta_int_lowest}
\end{equation}
Higher order diagrams can be obtained from \eqref{eq:delta_int_lowest} by including more fields. For instance, from the first term in Eq.~\eqref{eq:delta_int_lowest}, one can add any number of pairs $\Delta^{\dagger} \Delta$ between the original $\Delta$ and $R'$. The sum of all such diagrams is 
\begin{equation}
    \tr \left[ \Delta \left( \id + \frac{\Delta^{\dagger} \Delta}{k_E^2} + \frac{(\Delta^{\dagger} \Delta)^2}{k_E^4} + \cdots \right) R' \Delta^{\dagger} R \right] 
    \equiv  \tr \left[ \{\Delta\} R' \Delta^{\dagger} R  \right] \,,
\end{equation}
where in the second term, we have introduced the notation $\{\Delta\}$ for the infinite sum of $\Delta$. Since the cyclic symmetry between the $(\Delta^{\dagger}\Delta)$ vertex pairs is broken by the presence of other fields in the sequence, there is no symmetry factor accompanying each new diagram. The sum of these diagrams is thus $\{ \Delta \}$ instead of $\Delta \log (1 + |\Delta|^2/k_E^2)$, which would be the case if the other fields were not present. One can now include diagrams that repeat other fields, and the sum of all these diagrams is
\begin{equation}
    \tr \left[ \{\Delta\}  \{R'\}  \{\Delta^{\dagger}\}  \{R\} \right] \,.
    \label{eq:delta_int_repeat}
\end{equation}
Of course for the original sequence in Eq.~\eqref{eq:delta_int_lowest}, instead of contracting the last $R$ with the first $\Delta$ to form the loop, one can include a new $\Delta$ in the sequence, which must also include $R', \Delta^{\dagger}$ and $R$ at least once as dictated by gauge invariance. This is equivalent to repeating the sequence in Eq.~\eqref{eq:delta_int_repeat} any amount of times. The sum of these sequences is
\begin{equation}
    V_{\rm mix} \supset  \int \frac{d^4k_E}{(2\pi)^4} \tr \left[ 
        \log \left( 1 + \frac{ \{\Delta\}\{R'\}\{\Delta^{\dagger}\}\{R\} }{k_E^4} \right) 
        + \hc \right] \,,
    \label{eq:veff}
\end{equation}
where the log term now takes into account the cyclic symmetry. This would be the full one-loop effective potential if $R, R'$ are complex. Instead, there are additional terms such as
\begin{equation}
    \tr \left[ \Delta R'^2 \tDelta R^2 \right]\,,
\end{equation}
which have not been taken into account in Eq.~\eqref{eq:delta_int_repeat} since it contains only odd powers of $R$ and $R'$ (inside the logarithmic sum). These terms can easily be included by adding a new term $\tr \left[ \{\Delta\} R' \{R'\}  \{\Delta^{\dagger}\} R \{R\} \right]$, which is further suppressed by $1/\Lambda_{\rm UV}^2$. Since we are only interested in the leading order, these terms will be ignored when we derive the gap equation. Again, note that the effective potential in~\eqref{eq:veff} should only be used as a book keeping device to list out the diagrams. To obtain the exact one loop effective potential, one needs to perform a nontrivial summation of higher order diagrams. 

The integral over the lowest order~\eqref{eq:delta_int_lowest}, however, contains an infrared divergence, while there is no closed-form expression for the integral over~\eqref{eq:delta_int_repeat} and in~\eqref{eq:veff}. When deriving the gap equation for a particular field $\phi$, a simple trick to avoid the infrared divergence is to retain the corresponding sum $\{\phi\}$, while dropping higher order terms for the other fields. For instance, choosing $\Delta$ and keeping only $\{\Delta\}$ in \eqref{eq:delta_int_repeat}, the approximation for the mixing term is:
\begin{align}
    V_{\rm mix} &\supset \int \frac{d^4k_E}{(2\pi)^4} \frac{1}{k_E^4}  
        \tr \left( \frac{\Delta}{1 + \frac{ \Delta^{\dagger} \Delta }{k_E^2}} \tPhi^{\dagger} \tDelta \Phi^{\dagger} + \hc \right)\,.
\end{align}
The integral can now be performed, giving the effective potential in Eq.~\eqref{eq:delta_veff}.

\section{Hypercolor Nambu-Goldstone boson spectrum}
\label{app:B}

For the sake of completeness, we discuss the hypercolor Nambu-Goldstone bosons resulting from the enlarged color breaking. 
In the symplectic hypercolor case, the mass matrix in the $(\bm{\chi}, \chi_c, \chi', \bm{\bchi}, \bchi_c, \bchi')$ hyperfermion basis is
\begin{equation}
    \left( \begin{array}{@{}ccc|ccc@{}}
         & &                                                &  &  & \Delta_n \id_{N} \\
        \multicolumn{3}{c|}{ {\bf 0}_{ (2N + 3) } }  &  & m_{\chi} \id_{3} &  \\
         & &                                                & \tDelta_n \id_{N} &  &  \\ \hline
         &  & -\Delta_n \id_{N} & & &   \\
         & -m_{\chi} \id_{3} &  & \multicolumn{3}{c}{ {\bf 0}_{ (2N + 3) } }   \\
        -\tDelta_n \id_{N} &  &  & & &     
    \end{array} \right)  \,.
    \label{eq:gen_vacuum}
\end{equation}
Similar to the analysis in Section~\ref{sec:hc_enlarged_color}, the fermion condensates $\braket{\chi \bchi}, \braket{\chi'\bchi'}$ arising from the trace $\chi_F \bchi^F, \chi'_f, \bchi'^f$ are excluded. It is straightforward to see that including these terms simply shifts $m_{\chi}$ and rotates $\Delta, \tDelta$ among each other, and thus have no effect on the resulting Nambu-Goldstone boson structure. 
We have also excluded the subleading term $m_{\chi} \bm{\chi} \bm{\bchi}$ since we have taken $m_{\chi} \ll \Delta_n$. However, $m_{\chi}$ is the leading mass contribution to $\chi_c$, and thus has been retained in~\eqref{eq:gen_vacuum}. 
The VEV \eqref{eq:gen_vacuum} breaks the global symmetry
\begin{equation}
    G_{\Sp} = \U(4N + 6) \to \Sp(4N + 6) \,.
    \label{eq:sp_breaking_pattern}
\end{equation}
If the hyperfermion representation is real instead of pseudoreal, then the unbroken symmetry is $\SO(4N + 6)$. 
The pure QCD case, where $N = 0$, was considered in~\cite{Cacciapaglia:2015eqa}.
For $\SU(4)$ hypercolor, the same mass matrix can be expressed in a simpler form
\begin{equation}
    \left( \begin{array}{ccc} 
    \bchi' & \bchi_c & {\bm{\bchi}}
    \end{array}\right)
    \left( \begin{array}{@{}ccc@{}}
        \Delta_n \id_{N } &  & \\
        & m_{\chi} \id_{3 } & \\
        & & \tDelta_n \id_{N }
    \end{array} \right)  
    \left( \begin{array}{c} 
    \bm{\chi} \\ \chi_c \\ \chi'
    \end{array}\right)  \,.
\end{equation}
This leads to the global symmetry breaking pattern
\begin{equation}
    G_{\SU} = \U(2N + 3)_L \times \U(2N + 3)_R \to 
    \U(2N + 3)_V \,.
    \label{eq:su_breaking_pattern}
\end{equation}
For either hypercolor group choices, the global symmetry of the color sector has greatly increased, resulting in multiple extra $\U(1)$ group factors upon symmetry breaking. 
It is important to check the breaking pattern and identify the anomalous symmetries, since they might contribute to the $\theta$-terms of the strong sectors and spoil the strong CP solution. 
We start with the simpler case $G_{\SU}$, and later give the details for $G_{\Sp}$.  

Just like the quark masses responsible for breaking the global chiral symmetry, the mass term $m_{\chi}, \Delta_n$ and $\tDelta_n$ breaks $G_{\SU}$ to the diagonal subgroup.
However, unlike in QCD, $\Delta$ and $\tDelta$ are not bare masses, but rather dynamically generated by four-fermion couplings. 
Similar to the pion octet in QCD, the Nambu-Goldstone boson spectrum in this case fits into an adjoint multiplet of $\SU(2N + 3)_V$.
To label these states, we proceed to gauge the unbroken $\U(2N + 3)_V$ by $\SU(3)_c \times \U(1)_X$ and $\SU(N)_D$, where $\U(1)_X$ is the global symmetry that will mix with another unbroken $\U(1)$ from the global symmetry of $\psi\psi$ to form the hypercharge $\U(1)_Y$. 
The unbroken subgroup contains $\U(3) \times \U(N)_{\Delta} \times \U(N)_{\tDelta}$, corresponding to the vectorlike transformations of $\chi_c \bchi_c$, $\bm{\chi} \bchi'$, and $\chi' \bm{\bchi}$.  
The $\U(1)$ parts in these groups are respectively (up to normalization)
\begin{align}
    \U(1)^V_{\Delta}: \quad \text{diag} ( \id_N , {\bf 0}_3, {\bf 0}_N, {\bf 0}_N , {\bf 0}_3, -\id_N )\,, \label{eq:u1vdelta} \\
    \U(1)^V_{\chi_c}: \quad \text{diag} ( {\bf 0}_N, \id_3, {\bf 0}_N, {\bf 0}_N, -\id_3,{\bf 0}_N ) \,, \label{eq:u1vc} \\
    \U(1)^V_{\tDelta}: \quad \text{diag} (  {\bf 0}_N , {\bf 0}_3, \id_N,  -\id_N  , {\bf 0}_3, {\bf 0}_N) \label{eq:u1vtdelta} \,,
\end{align}
where these vector generators have been written in the hyperfermion basis of \eqref{eq:gen_vacuum} to make the connection to the $\Sp$ case clearer. 
Thus the $\SU(3)$ and the $\U(1)^V_{\chi_c}$ symmetry in~\eqref{eq:u1vc} can be identified with $\SU(3)_c$ and $\U(1)_X$. 
We gauge both $\SU(N)_{\Delta}$ and $\SU(N)_{\tDelta}$ with the same $\SU(N)_D$, meaning we only gauge part of the global symmetry corresponding to identical $\SU(N)_{\Delta}$, $\SU(N)_{\tDelta}$ transformations. 
Note that the diagonal color admits a $\U(1)_D$, similar to the $\U(1)_X$ of the $\SU(3)_c$. 
If $m_{\chi} \bm{\chi} \bm{\bchi}$ is taken into account, the other two $\U(1)^V_{\Delta, \tDelta}$ symmetries in~\eqref{eq:u1vdelta} and~\eqref{eq:u1vtdelta} are explicitly broken, leaving only the linear combination $\U(1)_D \equiv \U(1)^V_{\Delta} \times  \U(1)^V_{\tDelta}$.
The spectrum labeled under $\SU(3)_c\times\SU(N)_D\times \U(1)_X \times \U(1)_D$ is thus
\begin{align}
    (\rep{8}, \e)_{(0, 0)} + 2 (\rep{3}, \brep{N})_{(x, -z)} + 2 (\brep{3}, \rep{N})_{(-x, z)} 
    +\,4 (\e, \rep{Ad})_{(0, 0)} + 5 (\e, \e)_{(0,0)} \,.
    \label{eq:su_ngbs}
\end{align}
Some of these fields are eaten by the broken $\SU(N + 3) \times \SU(N)'$ gauge bosons. Specifically, one of the adjoint states $(\e, \rep{Ad})$, corresponding to $\bm{\chi} \bchi'$ with $N^2 - 1$ degrees of freedom (d.o.f), is eaten by $\frac{1}{\sqrt{2}}(G_N - G')^a_{\mu}$. The complex bifundamental $\chi_c \bchi'$ with $2 \times 3 N$ d.o.f is eaten by $G^b_{\mu}$, and one of the real singlets corresponding to $\tr \bm{\chi} \bchi'$ is eaten by $G_{1 \mu}$. In total, there are $N(N + 6)$ Nambu-Goldstone bosons eaten. 

It is nontrivial to derive the mass of these states, since there are multiple sources that can contribute, depending on which Nambu-Goldstone boson is considered. 
For instance, the charged states in~\eqref{eq:su_ngbs} receive two sources of mass from $m_{\chi}, \Delta,\tDelta$ (similar to the pion masses being proportional to the quark masses), and from color and diagonal color loop corrections (similar to photon loop corrections for the QCD pion masses). Thus, these masses should be $\sim \Lambda_{\rm HC}$. The other four singlets in~\eqref{eq:su_ngbs} correspond to a $\tr \chi' \bm{\bchi}$, which is also heavy due to the contribution from $\tDelta \sim \Lambda_{\rm HC}$, and the remaining three Nambu-Goldstone bosons are associated with the anomalous generators
\begin{align}
    \U(1)^A_{\bm{\chi}}: \quad \text{diag} ( \id_N , {\bf 0}_3, {\bf 0}_N, \id_N, {\bf 0}_3, {\bf 0}_N )\,,  \\
    \U(1)^A_{\chi_c}: \quad \text{diag} ( {\bf 0}_N, \id_3, {\bf 0}_N, {\bf 0}_N, \id_3,{\bf 0}_N ) \,,  \\
    \U(1)^A_{\chi'}: \quad\text{diag} ( {\bf 0}_N , {\bf 0}_3, \id_N , {\bf 0}_N, {\bf 0}_3, \id_N )\,. \label{eq:u1primed}
\end{align}
These generators, respectively, correspond to the axial transformations of $\bm{\chi} \bm{\bchi}$, $\chi_c \bchi_c$, and $\chi' \bchi'$. Note that the product of these symmetries is the overall anomalous $\U(1)$ of the $\U(4N + 6)$. The $\U(1)^A_{\bm{\chi}, \chi_c}$ symmetries are explicitly broken by a mass term $m_{\chi} \chi \bchi = m_{\chi} \bm{\chi} \bm{\bchi} + m_{\chi} \chi_c \bchi_c$, while the $\U(1)^A_{\chi'}$ is broken by $m_{\chi'} \chi' \bchi'$. Thus like the $\eta'_{\rm QCD}$ receiving mass from QCD instantons, these states can also receive mass from the enlarged color instantons, even though they don't contribute to the strong CP solution. 

The $\Sp$ hypercolor follows a similar story. Given that $\SU(2N + 3)_V \subset \Sp(4N + 6) \subset \SU(4N + 6)$, we can decompose the Nambu-Goldstone bosons from the spontaneous breaking $\SU(4N + 6) \to \Sp(4N + 6)$ as $\rep{Ad} + \rep{A}_2 + \brep{A}_2$ under $\SU(2N + 3)_V$. Thus the spectrum contains all of the states in the $G_{\SU}$ case, namely the $\rep{Ad}$ multiplet, together with the additional multiplets $\rep{A}_2$ and $\brep{A}_2$ of $\SU(2N+3)_V$. After weakly gauging the global symmetry, $\SU(2N + 3)_V$ with $\SU(3)_c \times \SU(N)_D$, we obtain in addition to \eqref{eq:su_ngbs}, the spectrum 
\begin{align}
	(\rep{3}, \e)_{(2x, 0)} + (\brep{3}, \e)_{(-2x, 0)} + 2\,(\rep{3}, \rep{N})_{(x, z)} &+ 2\,(\brep{3}, \brep{N})_{(-x, -z)}    \notag \\
	+\, 4(1, \rep{A}_2)_{(0, 2z)} + 4\,(1, \brep{A}_2)_{(0, -2z)} + 2\,(1, \rep{S}_2)_{(0, 2z)} &+ 2\,(1, \brep{S}_2)_{(0, -2z)} + (\e, \e)_{(0, \pm 2z)} \,,
	\label{eq:sp_ngbs}
\end{align}
where the subscripts refer to the $\U(1)_X$, $\U(1)_D$ charges.
The two new singlets in $\eqref{eq:sp_ngbs}$ correspond to $\tr \bm{\chi} \chi'$ and $\tr \bm{\bchi} \bchi'$, which are obviously broken explicitly by $m_{\chi, \chi'}$. The detection of the states in~\eqref{eq:sp_ngbs} would be a way to distinguish the hypercolor gauge group.

In summary, the pseudo Nambu-Goldstone spectrum arising from the hypercolor top partners obtain masses depending on $\Lambda_{\rm HC}$ and $m_{\chi, \chi'}$. It can be expected that all of these states lie between the $\Lambda_D$ and $\Lambda_{\rm HC}$ scales, although the exact spectrum would require a more in-depth analysis.
Interestingly, as some of the hypercolor Nambu-Goldstone boson masses are proportional to $\sqrt{m_{\chi, \chi'} \Lambda_{\rm HC}}$, these bound states could be experimentally accessible if $m_{\chi, \chi'}\ll \Lambda_{\rm HC}$.
Finally, in both hypercolor cases, there are unbroken $\U(1)$ groups, which can be associated with the baryon number of the QCD and $\SU(N)_D$ quarks. It is possible, though unnecessary for our purposes, to define a generalized baryon number that combines the $\U(1)_D$ and QCD baryon numbers. Alternatively, one can gauge the $\U(1)_D$ symmetry to give a dark photon, which could be a potential dark matter candidate.

\newpage

\newpage
\bibliographystyle{JHEP}
\bibliography{citations}

\end{document}